\documentclass[prl,twocolumn,preprintnumbers,superscriptaddress,longbibliography]{revtex4-1}

\usepackage{graphicx}
\usepackage{bm}
\usepackage{amsmath}
\usepackage{amssymb}
\usepackage{amsfonts}
\usepackage{fancyhdr}
\usepackage{latexsym,epsfig,bbm}
\usepackage{algorithm}
\usepackage{algorithmic}
\usepackage{amsthm}

\usepackage{color}
\definecolor{blue}{rgb}{0,0.2,1}

\definecolor{red}{rgb}{0.9,0,0}

\newcommand{\hdef}[0]{h_{\mathrm{dflt}}}
\newcommand{\Ydef}[0]{Y_{\mathrm{dflt}}}
\newcommand{\rhodef}[0]{\rho_{\mathrm{dflt}}}
\newcommand{\init}[0]{\mathrm{init}}
\newcommand{\mg}[1]{#1}

\newcommand{\ket}[1]{|#1\rangle}
\newcommand{\bra}[1]{\langle #1 |}

\newtheorem{definition}{Definition}
\newtheorem{result}{Result}
\newtheorem{theorem}{Theorem}

\newtheoremstyle{named}{}{}{\itshape}{}{\bfseries}{.}{.5em}{\thmnote{#3's }#1}
\newtheoremstyle{thm}{}{}{\itshape}{}{\bfseries}{.}{.5em}{#1 \thmnote{(#3)}}
\theoremstyle{named}

\theoremstyle{thm}
\newtheorem{thm}{Theorem}

\makeatletter
\DeclareRobustCommand{\cev}[1]{%
  \mathpalette\do@cev{#1}%
}
\newcommand{\do@cev}[2]{%
  \fix@cev{#1}{+}%
  \reflectbox{$\m@th#1\vec{\reflectbox{$\fix@cev{#1}{-}\m@th#1#2\fix@cev{#1}{+}$}}$}%
  \fix@cev{#1}{-}%
}
\newcommand{\fix@cev}[2]{%
  \ifx#1\displaystyle
    \mkern#23mu
  \else
    \ifx#1\textstyle
      \mkern#23mu
    \else
      \ifx#1\scriptstyle
        \mkern#22mu
      \else
        \mkern#22mu
      \fi
    \fi
  \fi
}

\DeclareMathAlphabet\mathbfcal{OMS}{cmsy}{b}{n}
\newcommand{\past}[1]{\cev{#1}}
\newcommand{\future}[1]{\vec{#1}}
\newcommand{\omni}[1]{\overleftrightarrow{#1}}

\newcommand{\arxiv}[1]{}

\newcommand{\dflt}{{\rm dflt}}

\begin{document}




\title{Energetic advantages for quantum agents in online execution of complex strategies}

\author{Jayne Thompson}
\email{thompson.jayne2@gmail.com}
\affiliation{ Institute of High Performance Computing, Agency for Science, Technology and Research (A*STAR), Singapore}

\author{Paul M.~Riechers}
\email{pmriechers@ucdavis.edu}
\affiliation{Beyond Institute for Theoretical Science (BITS), San Francisco, CA}

\author{Andrew J.~P.~Garner}
\email{ajp.garner@gmail.com}
\affiliation{Institute for Quantum Optics and Quantum Information, Austrian Academy of Sciences, Boltzmanngasse 3, 1090, Vienna, Austria}

\author{Thomas J.~Elliott}
\email{physics@tjelliott.net}
\affiliation{Department of Physics \& Astronomy, University of Manchester, Manchester M13 9PL, United Kingdom}
\affiliation{Department of Mathematics, University of Manchester, Manchester M13 9PL, United Kingdom}

\author{Mile Gu}
\email{mgu@quantumcomplexity.org}
\affiliation{Nanyang Quantum Hub, School of Physical and Mathematical Sciences, Nanyang Technological University, 637371, Singapore}
\affiliation{Centre for Quantum Technologies, National University of Singapore, 3 Science Drive 2, 117543, Singapore}

\begin{abstract}
Agents often execute complex strategies -- adapting their response to each input stimulus depending on past observations and actions. Here, we derive the minimal energetic cost for classical agents to execute a given strategy, highlighting that they must dissipate a certain amount of heat with each decision beyond Landauer's limit. We then prove that quantum agents can reduce this dissipation below classical limits. We establish the necessary and sufficient conditions for a strategy to guarantee quantum agents have energetic advantage, and illustrate settings where this advantage grows without bound. Our results establish a fundamental energetic advantage for agents utilizing quantum processing to enact complex adaptive behaviour.
\end{abstract}

\maketitle

A blackjack player counting cards, a control system monitoring a production line, autonomous vehicles navigating busy streets -- all represent examples of online agents executing adaptive strategies. Online, in that they must decide each response without foreknowledge of future input~\cite{albers2003online}; and adaptive in that optimal output behaviour depends not only on the present stimuli but also on past events ~\cite{barnett2015computational}. As we automate complex tasks of ever-growing complexity, the resulting energetic costs grow unsustainably~\cite{thompson2020computational,strubell2019energy}, 
presenting an ultimate performance bottleneck and necessitating performance-power trade-offs~\cite{mcdonald2022great}. 

\begin{figure}[!t]
	\centering
	\includegraphics[width=
0.833\linewidth]{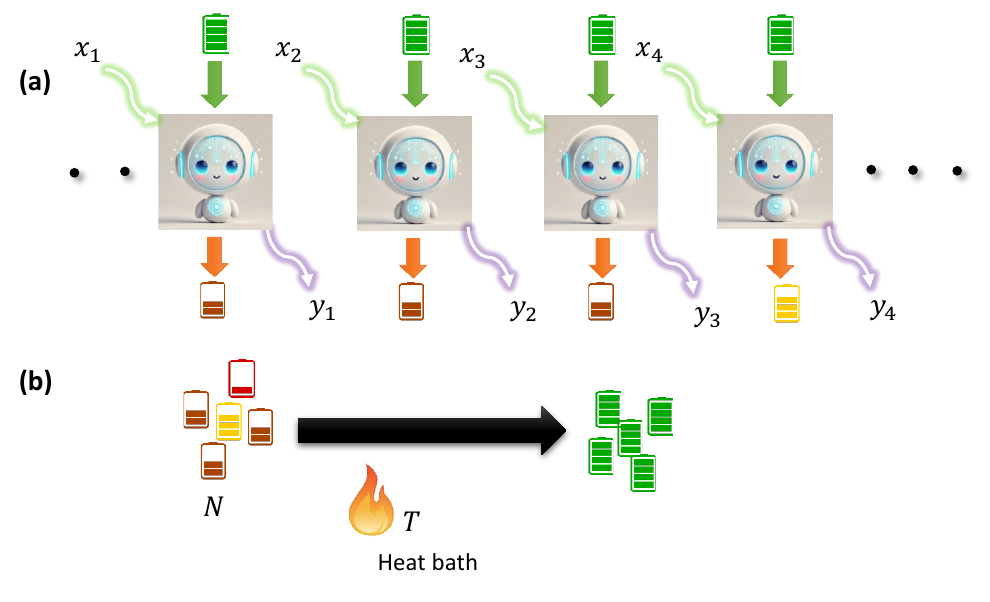}\caption{\label{figframework} \textbf{Agent energetics} (a) At each time-step $t$, the agent receives an input question $x_t$ and provides an answer $y_t$ according to some desired strategy $\mathcal{P}$. This process requires energy, which is drawn from an information battery. At the end of each time step the depleted battery is replaced by a fresh battery, after which the agent can respond to the next input $x_{t+1}$. (b) Afterwards we can take an ensemble of depleted batteries (collected from $n$ agents operating in parallel) and collectively reset them while in contact with a heat bath at temperature $T$. This step costs $E_n$ units of energy. The energetic cost of each agent is then $w = E_n/n$ in the thermodynamic limit of $n \gg 1$. }
 \end{figure}

Does physics place fundamental limits on energy expenditure for executing a complex adaptive strategy online? If so, can quantum agents operate at energy efficiencies that are classically unreachable? Here, we introduce a framework to rigorously quantify an agent's energetic costs (see Fig.~\ref{figframework}), and derive a fundamental bound on the minimal energy requirements of a classical agent executing any designated strategy. We then isolate necessary and sufficient conditions for a strategy to be executable by a quantum agent with reduced energy dissipation and illustrate scenarios where this advantage can grow without bound. These energetic advantages do not require the agent to receive inputs or emit outputs in quantum superposition and thus persist when quantum agents interact with purely classical environments. Thus, we identify a new form of quantum advantage applicable in all situations where classical agents are used.

\textbf{Framework} -- We formalize complex strategies by considering a two-party game between an agent and an interrogator over discrete time-steps $t \in \mathbb{Z}$. At each time-step $t$, the interrogator sends an input query $x_t \in \mathcal{X}$, requiring the agent to respond with some output $y_t \in \mathcal{Y}$. 
We describe the input-output pair by $z_t := (x_t,y_t)$. We define the past history of inputs and outputs as $\past{z} := \ldots z_{-2}z_{-1}$, such that the agent is currently waiting for input query $x_0$ at $t = 0$. Thus, we can denote future input-outputs by $\future{z} :=  z_{0}z_{1}z_{2} \ldots$.

A strategy $\mathcal{P}$ describes desired input-output behaviour~\cite{barnett2015computational,crutchfield1989inferring}.  Each strategy $\mathcal{P} = \{P(Y_{0:K} = y_{0:K}| x_{0:K}, \past{z})\}_{K > 0}$ specifies the probability with which the agent outputs $y_{0:K} = y_0y_1\ldots y_{K-1}$ when given a sequence of $K$ future inputs $x_{0:K} = x_0x_1\ldots x_{K-1}$ for each natural number $K$, conditioned on history $\past{z}$ \footnote{Note that for each potential future input string $\future{x} = x_0x_1\ldots$  we require there to be a valid stochastic process satisfying the Kolmogorov extension theorem such that for each finite length $K$ we can recover $P(Y_{0:K} = y_{0:K}| x_{0:K}, \past{z})$ as a valid marginal distribution.}. Note that while this definition specifies the random variable $Y_t$ that governs each $y_t$, it makes no such specification on $x_t$. A strategy must describe the agent's response for every possible future input $x_t$, regardless of how $x_t$ is distributed. We thus adopt similar conventions to Bell tests, such that the agent gains no information about future inputs based on past inputs. Therefore, agents minimizing heat dissipation are best placed assuming each $x_t$ to be independent and identically distributed (with Shannon entropy $h_x$)~\cite{kolchinsky2017dependence,Riechers2021initial}. Here we are interested in the average cost per time step (see Fig.~\ref{figframework}) and focus on \emph{stationary} (i.e., time-translation invariant) strategies where $P(Y_{t:t+K} = y_{t:t+K}|x_{t:t+K}, \past{z}_t)$ has no explicit dependence on $t$.

\begin{figure}[!t]
	\centering
	\includegraphics[width=
1.05\linewidth]{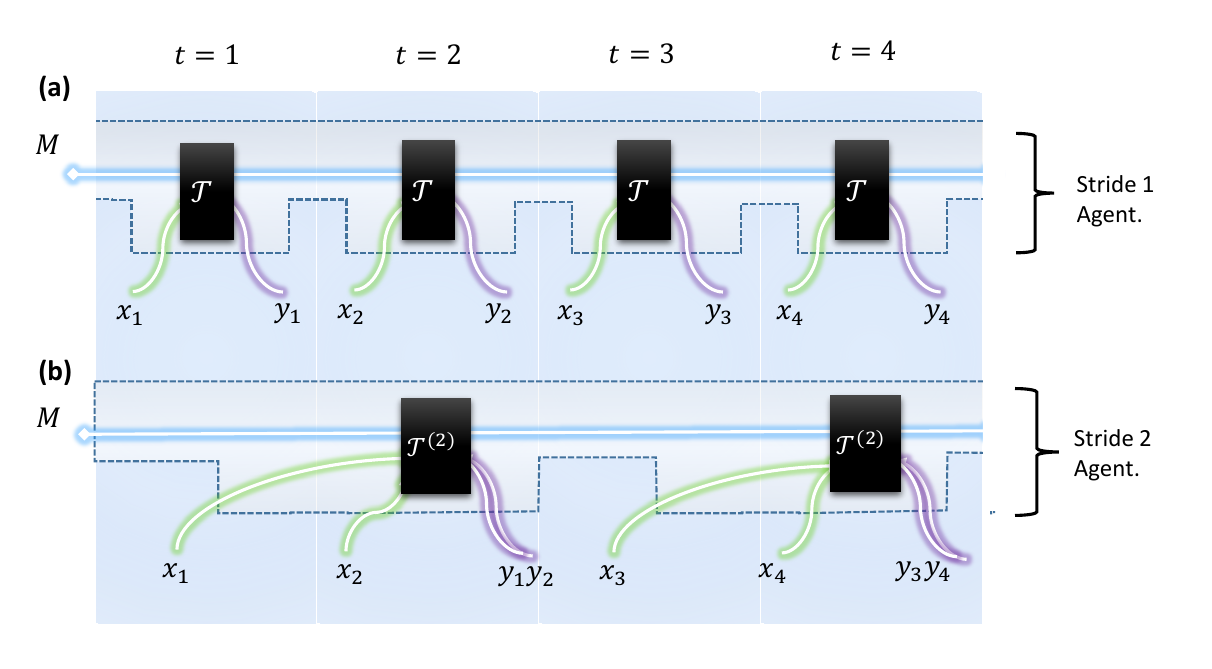}\caption{\label{figstride} \textbf{Agents and strides.} (a) The action of an online agent ($L = 1$) can be represented as a circuit. Here repeatedly applying the policy map $\mathcal{T}$, repeatedly couples the memory $\mathbf{M}$  with inputs $x_t$, to emit outputs $y_t$.  To study the energetic cost of online response. We also consider $L$-stride agents that process $L$ inputs at a time, such as in (b) where $L = 2$.}
 \end{figure}

Unrestricted, an agent can withhold committing an output until an arbitrary number of future questions are asked. Here, we are interested in agents operating \emph{online}: they must emit a particular $y_t$ before input $x_{t+1}$ is given (See Fig.~\ref{figstride}a).  To understand the resulting thermodynamic costs, we introduce quasi-online $L$-stride agents that must commit outputs every $L$ time-steps (see Fig.~\ref{figstride}b). Using only $\mathbf{M}$, the $L$-stride agent must output statistically appropriate $y_{0:L}$ upon receipt of any future inputs $x_{0:L}$, while updating their memory to enable correct generation of $y_{L:2L}$ when given $x_{L:2L}$.  All such agents must host a memory $\mathbf{M}$ along with an encoding function $f:\past{\mathcal{Z}}\to\mathcal{R}$, that configures the memory in a state  $r = f(\past{z}) \in \mathcal{R}$ containing all relevant information about $\past{z}$. We assume all agents are causal such that $\mathbf{M}$ only encodes information about the future already contained in $\past{z}$~\cite{elliott2022quantum}.

Formally, such agents operate via a systematic map $\mathcal{T}^{(L)}$ on $\mathbf{M}$ and input $x_{0:L}$ that (i) emits $y_{0:L}$ sampled from $P(Y_{0:L} | x_{0:L}, \past{z})$ and (ii) updates their memory from $r_0 = f(\past{z})$ to $r_{L} = f(\past{z}')$, where $\past{z}' = (\past{x} x_{0:L}, \past{y}y_{0:L})$. The memory states $\mathcal{R} = \{r_1,r_2,\ldots, r_m\}$ are often \mg{named} \emph{belief states}, representing an agent's belief about the present based on past experience~\cite{marzen2018optimized,zhang2019learning}. Meanwhile, $\mathcal{T}^{(L)}$ describes the agent's \emph{policy} - their mechanism for choosing the next output based on present input and belief~\cite{sutton1999reinforcement}. Every $L$-stride agent can be described by the tuple $(\mathcal{X}, \mathcal{Y}, \mathcal{R}, f, \mathcal{T}^{(L)})$. The online ($L=1$) case aligns with previous definitions of agents and information transducers~\cite{barnett2015computational}. The case of large $L$ corresponds to sequence-to-sequence generators~\cite{sutskever2014sequence}.

\textbf{Agent Energetics} -- 
We need a detailed description of the agent's internal mechanics to determine its energetic cost. Let $\mathbf{X}_t$ and $\mathbf{Y}_t$ denote the physical systems that respectively encode $x_t$ and $y_t$. Thus $\mathbf{X}_{0:L} = \mathbf{X}_0,\mathbf{X}_1\ldots,\mathbf{X}_{L-1}$ and $\mathbf{Y}_{0:L} = \mathbf{Y}_0,\mathbf{Y}_1\ldots,\mathbf{Y}_{L-1}$ represent a physical tape of $L$ such systems that respectively encode $L$ consecutive inputs and outputs. Before interacting with the agent, the interrogator sets each $\mathbf{X}_t$ to the appropriate question $x_t$, and all $y_t$ are initially maximally mixed with entropy $\hdef = \log_2 |\mathcal{Y}|$~\footnote{This assumption ensures that the tape for recording outputs is not a source of free-energy, and all thermodynamic resources injected are accounted for.}. An agent imprints its actions on the output tape, transforming $\mathbf{Y}_t$ to encode $y_t$ with probabilities dictated by the target strategy $\mathcal{P}$. We assume the Hamiltonians for the information tapes are fully degenerate at the start and end of the protocol, and that the encodings for inputs and outputs are classical~\footnote{That is, $x_t$ and $y_t$ are encoded in mutually distinguishable states $\ket{x_t}$ and $\ket{y_t}$. That is
if $x_t \neq x'_t$ then $\langle x_t | x'_t \rangle = 0$}. This is true regardless of whether we employ classical or quantum agents, ensuring they play by the same rules \footnote{Note that this assumption distinguishes our results from quantum active learning agents which derive quantum advantage through coherent interactions with their environment \cite{paparo2014quantum,saggio2021experimental}.}.

The agent's policy map $\mathcal{T}^{(L)}$ is then some physical process that transforms the input tape $\mathbf{X}_{0:L}$, a tape of $L$ maximally mixed states $\mathbf{Y}_{0:L}$, and its memory $\mathbf{M}$ initially in $f(\past{z})$, such that after application: (a) $\mathbf{Y}_{0:L}$ encodes $y_{0:L}$ with probability \mbox{$P(Y_{0:L} = y_{0:L}| x_{0:L}, \past{z})$};
 (b) $\mathbf{M}$ encodes $f(\past{x} x_{0:L}, \past{y} y_{0:L})$; and (c) the state of $\mathbf{X}_{0:L}$ is unchanged. Conditions (a) and (b) guarantee that the agent faithfully executes the strategy $\mathcal{P}$. (c) ensures the agent is not cannibalising $x_{0:L}$ as a source of free energy~\footnote{Note that this assumption distinguishes our results from that of information-ratchets -- which focus on extracting free-energy from input sequences~\cite{mandal2012work, boyd2016identifying}.}. 

\arxiv{\begin{enumerate}
\setlength{\itemsep}{0pt}
    \item[(a)]  $\mathbf{Y}_{0:L}$ encodes $y_{0:L}$ with probability \mbox{$P(Y_{0:L} = y_{0:L}| x_{0:L}, \past{z})$}.
    \item[(b)]  $\mathbf{M}$ encodes $f(\past{x} x_{0:L}, \past{y} y_{0:L})$.
    \item[(c)]  The state of $\mathbf{X}_{0:L}$ is unchanged.
\end{enumerate}}

Denote the energetic cost of executing $\mathcal{T}^{(L)}$ by $W^{(L)}$. An $L$-stride agent with policy $\mathcal{T}^{(L)}$ would thus require $W^{(L)}$ units of work to generate $L$ sequential output responses - and thus have \emph{work rate} (work cost per time-step) of $w^{(L)} = W^{(L)}/L$. $\mathcal{T}^{(L)}$ is generally compressive, with initial entropy $H_i = L \hdef + L h_x + H(M_0)$ and final entropy $H_f = H(Z_{0:L}, M_{L})$, where $M_t$ is the random variable governing the state of $\mathbf{M}$ at time $t$. Setting $k$ as Boltzmann's constant and $T$ as the temperature of the thermal reservoir, Landauer's principle then implies (see Supplementary Material C for details):

\begin{result}\label{result:1}
The work rate of any agent, classical or quantum, is bounded from below by
\begin{equation}\label{eq:optimalratchet}
\frac{w^{(L)}}{kT \ln 2} \geq \hdef + \frac{1}{L}[I(Z_{0:L};M_L) - H (Y_{0:L}|X_{0:L})],
\end{equation}
where $I(A;B) = H(A) + H(B) - H(A,B)$ denotes the mutual information, and $H(A|B) = H(A,B) - H(B)$ is the conditional entropy. \end{result}

\textbf{Optimal classical agents} -- Classical agents have classical memory, and can saturate the above bounds using isothermal channels and changing energy landscapes~\cite{boyd2018thermodynamics,loomis}.
Therefore, the energy-minimal agent should choose a memory encoding $f(\past{z})$ that minimises $I(Z_{0:L};M_L)$. This minimum is attained when an agent allocates memory to distinguish two pasts iff their required future statistical responses differ~\footnote{We briefly review these concepts in Supplementary Materials B. For further information, see ~\cite{barnett2015computational,shalizi2001computational}.}. Mathematically, the encoding function of such agents satisfy $\epsilon(\past{z}) = \epsilon(\past{z}')$ iff $P(Y_{0:K}| x_{0:K}, \past{z}) = P(Y_{0:K}| x_{0:K}, \past{z}')$ for all potential future input sequences  $x_{0:K}$ and all $K$. The resulting belief states $\mathcal{S} = \{s_1,\ldots, s_m\}$ are known as the \emph{causal states} of the target strategy $\mathcal{P}$~\cite{barnett2015computational,zhang2019learning}. 

The causal states induce a family of energetically-minimal agents for each stride $L$. When $L = 1$, the associated online agent is known as the \emph{$\epsilon$-transducer}~\cite{barnett2015computational}. We can represent its policy $\mathcal{T}^{(1)}$ by a collection of stochastic maps $T^{y|x}_{jk}$ -- the probability a memory initially in state $s_j$ transitions to $s_k$ while outputting $y$, conditioned on receiving input $x$. Concatenation of this map over $L$ time-steps defines the policy map of the associated $L$-stride agent -- specified by $T^{y_{0:L}|x_{0:L}}_{jk}$, the probability the machine will output $y_{0:L}$ over the next $L$ time-steps on input $x_{0:L}$ while transitioning from $s_j$ to $s_k$. \mg{Let $S_L$ be the random variable governing the causal state of $\mathcal{P}$ after application of $\mathcal{T}^{(L)}$. 
 The minimal work cost for any classical agent is then given by Result \ref{result:1} with $M_L = S_L$}.

\textbf{Extra work cost of online response} -- The results above indicate that classical agents incur a fundamental energetic cost to respond online. As the stride length \( L \rightarrow \infty\): 
$$ \frac{w^{(L)}_c}{kT \ln 2} \rightarrow \hdef - \frac{1}{L}H(Y_{0:L}|X_{0:L}).$$
The optimal work rate in the limit that the agent has no online response constraints thus aligns with the change in free energy of the tape. Such agents can saturate Landauer's limit, thus operating reversibly and dissipating no heat. The difference 
\begin{equation}
w_{\mathrm{onl}} =  w^{(1)}_c - \lim_{L \rightarrow \infty} w^{(L)}_c,
\end{equation}
between this quantity and energy cost of an online agent represents the \emph{work cost of online response} -- the extra dissipative work cost for an agent to operate online. In Supplementary Materials F, we show that for optimal classical agents, the minimal extra dissipation is
\begin{eqnarray}\label{eqn:erase}
w_{\mathrm{onl}} &=& kT \ln 2[ I(Z_0;S_1) - I(Z_1; S_1)].
\end{eqnarray}
\arxiv{Online response places additional constraints on the agent, forcing tracking and erasure of otherwise unnecessary information.}
This extra heat dissipation remains even when we saturate Landauer's limit at each time-step \footnote{This overhead is reminiscent of modularity dissipation ~\cite{kolchinsky2017dependence, boyd2018thermodynamics}, where we pay an additional cost for time-local information processing~\cite{loomis2020thermodynamically, elliott2021memory}}. An online agent lacks foreknowledge of future inputs, and thus is forced to optimise thermal efficiencies of the computation piecemeal. \arxiv{Achieving thermodynamic reversibility of individual logical gates in a circuit does not imply achieving thermodynamic reversibility in the entire circuit.}

\begin{figure}
	\centering
	\includegraphics[width=
1\linewidth]{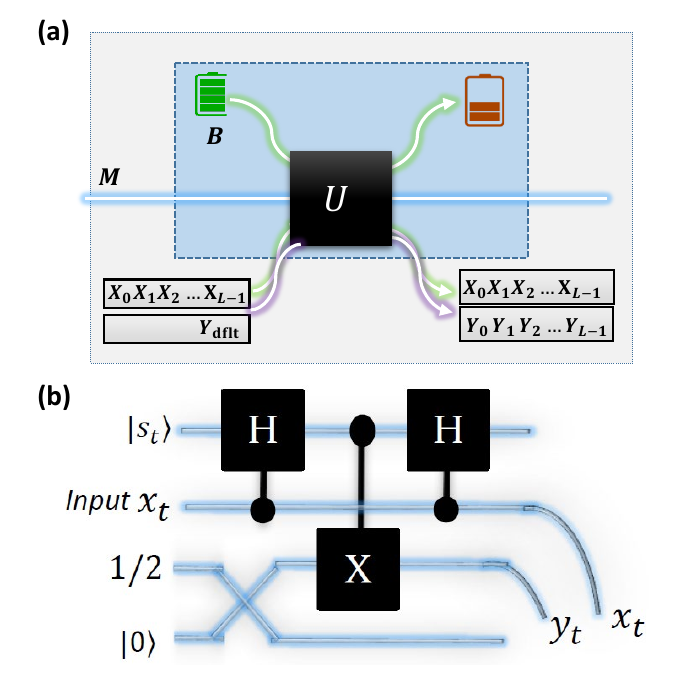}
	\caption{\label{figmarkovblank} \textbf{Information Batteries.} (a) The energetic cost of a quantum agent executing a policy map $\mathcal{T}^{(L)}$ can be characterised using an information battery $\mathbf{B}$. Here $U$ represents a Stinespring dilation of $\mathcal{T}^{(L)}$ that couples $\mathbf{B}$ with the agent's memory and the input and output tapes $\mathbf{X}_{t:t+L}$  and $\mathbf{Y}_{t:t+L}$. At the end of the operation, the depleted battery is ejected. The energy cost needed to reset this battery when optimised over all possible Stinespring dilations gives the single-shot work cost in implementing $\mathcal{T}^{(L)}$, and can be directly computed \cite{delrio, faist, vitanov,  tomamichel2015quantum, fawzi2015quantum,renner2008security, junge2018universal,tomamichel2010duality}. In the i.i.d.~limit, jointly resetting these batteries incurs a work rate of $w^{(L)}_q$. (b) Quantum Alice's circuit in our example where the gates are controlled unitaries $C-V = \ket{0}\bra{0} \otimes I + \ket{1}\bra{1}\otimes V$ such that $I$ is the identity, $H = \ket{+}\bra{0} + \ket{-}\bra{1}$ is the Hadamard gate, and  $X = \ket{1}\bra{0} + \ket{0}\bra{1}$ is the Pauli-$X$ gate. Similarly $I/2 = 1/2(\ket{0}\bra{0} + \ket{1}\bra{1})$ is the completely mixed state. Here Alice requires $kT\ln 2$ units of work to reset the depleted battery register (bottom wire).}
 \end{figure}
 
\textbf{Quantum agents} -- Quantum agents utilize quantum memory~\cite{elliott2022quantum}, allowing each causal state \( s_k \) to be associated with a quantum memory state \( |\sigma_k\rangle \). Formally, consider the encoding function $\epsilon_q  = \psi_q \, \circ \, \epsilon$ where $\epsilon$ is the classical encoding map onto causal states and $\psi_q : \mathcal{S} \rightarrow \{\ket{\sigma_j}\bra{\sigma_j}\}$ maps each causal state $s_j$ to an associated quantum memory state $\ket{\sigma_j}$.

We analyse the work cost of this operation using information batteries~\cite{faist, delrio}, mirroring techniques used for quantum simulators~\cite{loomis}. The approach involves an ancillary information battery  composed of $\lambda_1 \gg 1$ pure qubits and $\lambda_2 \gg 1$ maximally mixed qubits
(see Fig.~\ref{figmarkovblank}). Once configured in the appropriate memory state $\epsilon_q(\past{z}) = \ket{\sigma_j}\bra{\sigma_j}$, an $L$-stride quantum agent operates by applying  Stinespring dilation $U$ of $\mathcal{T}^{(L)}$ jointly on: (1) the agent's memory $\mathbf{M}$; (2) input system $\mathbf{X}_{0:L}$; (3) output system $\mathbf{Y}_{0:L}$; and (4)   a subsystem of qubits in the battery  $\mathbf{B}$. Through coupling to the battery, $U$ must map the initial state of input tape, output tape and memory $\ket{x_{0:L}}\ket{y_{\init_{0:L}}}\ket{\sigma_j} $ \mg{(where $\ket{y_{\init_{0:L}}}$ is an arbitrary initial state of the output tape)} to a superposition state
\begin{equation}\nonumber
\sum_{y_{0:L}, k}\sqrt{T^{y_{0:L}|x_{0:L}}_{j k}} \ket{z_{0:L}}_{Z_{0:L}}\ket{\sigma_{k}}_M\ket{\psi(z_{0:L},j, y_{\init_{0:L}})}_B
\end{equation}
where  $\ket{\psi(z_{0:L},j,y_{\init_{0:L}})}$ are junk states accumulated on the battery. Unitaries satisfying the above conditions can be systematically found for any given strategy (see Supplementary Materials B for details \cite{thompson2017using, elliott2022quantum}). Repeated action of any such $U$ enables an $L$-stride execution of strategy $\mathcal{P}$. This process changes the battery register $\mathbf{B}$. The minimal work cost for implementing $\mathcal{T}^{(L)}$ then corresponds to the energy required to reset $\mathbf{B}$ back to its initial state (assuming the optimal choice of $U$ and \mg{no pre-knowledge of $\ket{y_{\init_{0:L}}}$})~\cite{faist}. We note that in this picture, 
minimising heat dissipation requires only the batteries to be reset quasi-statically. The quantum agents themselves can execute a desired strategy in real time without necessarily sacrificing energetic efficiency.

Using techniques pioneered for resetting a system conditioned on a quantum memory \cite{delrio}, we upper bound the work cost of realising such an operation in the single-shot setting subject to a fidelity $\varepsilon$ and with failure probability at most $\delta$ (see Supplementary Materials D). We then take the i.i.d.~limit, corresponding to the per battery cost of simultaneously resetting a large number of such depleted batteries. The resulting work rate $w^{(L)}_q$ can saturate Eq.~(\ref{eq:optimalratchet}). Thus, the energetic advantage (per time-step) of a quantum agent over a classical agent is
\begin{equation}
w^{(L)}_c - w^{(L)}_q = \frac{kT \ln 2}{L}[I(Z_{0:L}; S_L) - I(Z_{0:L}; M_L)],
\end{equation}
where $I(Z_{0:L}; M_L)$ represents the amount of information our quantum agent retains about the past $L$ input-output pairs. \arxiv{Note that this i.i.d.~limit typically can only be approached by resetting many batteries concurrently  (see protocol in Fig.~\ref{figframework}).} Quantum agents are thus more energetically efficient if $I(Z_{0:L}; M_L)$ is lower than the minimal classical counterpart $I(Z_{0:L}; S_L)$. 

To determine conditions for quantum energetic advantage (i.e., $w^{(L)}_q < w^{(L)}_c$), consider an interrogator challenged to determine whether an $\epsilon$-transducer is in one of two possible causal states $s_j$ or $s_k$ at $t=0$. They cannot directly access the transducer's internal state but can adopt any interrogation strategy $\Lambda$. i.e., they can freely decide which $x_t$ to ask the transducer at each time-step $t$, resulting in a sequence of transducer outputs $y_t$ governed by either $P_{\Lambda}(\future{Y} | s_j )$ or $P_{\Lambda}(\future{Y}|s_k)$. Such interrogation strategies are general input-output processes (see Supplementary Materials for details), allowing the interrogator to decide $x_t$ adaptively based on all past observations. 
Can such an interrogator succeed with certainty? If not, then it suggests the $\epsilon$-transducer has causal waste -- some information it stores in memory to distinguish $s_j$ and $s_k$ is never exhibited in future statistics. We say $s_j, s_k$ forms a causally wasteful pair. This leads to \emph{necessary and sufficient} conditions for quantum agents to have energetic advantage (proof in Supplementary Materials E):

\begin{result}\label{result:2}
A $L$-stride quantum agent can execute a given strategy $\mathcal{P}$ with strictly lower work cost than any classical counterpart, i.e.,
\begin{equation}
w^{(L)}_q < w^{(L)}_c
\end{equation}
 if and only if $\mathcal{P}$ has two causal states $s_j, s_k$ that form a causally wasteful pair and $P(s_j) \neq P(s_j|z_{0:L})$ for some string $z_{0:L}$ of $L$ inputs and outputs. Furthermore, the memory states and policy of this quantum agent can be systematically constructed.
\end{result}

Combining the above results with the observation that  $w^{(L)}_q = w^{(L)}_c$ in limit of large $L$, we see that $w_c^{(1)} - w_q^{(1)}$ exactly measures the \emph{quantum energetic advantage in online response}. Furthermore, this is non-zero whenever optimal classical agents exhibit causal waste.

\textbf{Example} -- \mg{We illustrate above ideas via thought experiment. Consider an agent, Alice, under interrogation by Bob. At each time step Bob asks one of two binary questions at random ``Are you hungry?" ($x=0$) or ``Do you like sheep?" ($x=1$). If Bob repeats the same question in two consecutive time-steps, Alice's answers must agree; otherwise, her response to the second question must be random.}

Any classical agent must have 4 memory states -- aligning with the 4 possible question-answer pairs in the last time-step. Thus $h_{\dflt} =1$, $I(Z_0;S_1) = 2$ and $H(Y_0|X_0) = 1$. The resulting work rate is $w_c^{(1)} = 2 kT\ln 2$. Meanwhile, the work cost of online response is $w_{\mathrm{onl}} = 1.5 kT\ln 2$, as our agent stores $2$ bits about the immediate past, but these $2$ bits contain only $0.5$ bits about the future. Indeed this analysis corroborates studies of heat dissipation for certain realist interpretations of quantum mechanics~\cite{cabello2016thermodynamical}.

In contrast, a quantum agent can encode all $4$ causal states in a single memory qubit $\mathbf{M}$ by use of quantum belief states $\ket{0},\ket{+},\ket{1},\ket{-}$. The circuit in Figure \ref{figmarkovblank} (b) then generates desired input-output behaviour. Such a quantum agent would have $I(Z_0; M_1) = 1$,  and thus expends $kT\ln 2$ less energy per time-step, while heat dissipation (work cost of online response) is reduced is reduced from $1.5kT \ln 2$ to $0.5kT \ln 2$.

\textbf{Scaling Advantages} -- We highlight the potential for the gap between quantum and classical thermodynamic performance to scale in the Supplementary Materials H. In particular we give two processes which display a scaling advantage. One is based on a particle undergoing Brownian motion on a ring with sudden jumps upon input $x = 1$. Here the gap between the classical and quantum work cost diverges as we track the particle's position to higher and higher precision.
A second is based on the case where an agent makes decisions at discrete intervals seperated by $\tau$ seconds, but receives inputs every $\Delta t$ seconds. Such quasi-online agents essentially perform $L$-stride executions $\mathcal{P}_{\Delta t}$, with $L = \tau/\Delta t$. We outline a family of processes $\{\mathcal{P}_{\Delta t}\}$ where the dissipated work cost per unit time for classical agents then grows without bound as $\Delta t \rightarrow 0$. In contrast, quantum agents dissipated a bounded amount of energy even as $\Delta t$ approaches 0.

\textbf{Discussion} -- Complex adaptive strategies appear in diverse contexts, from navigating partially observable environments, to modelling non-Markovian noise and natural-language processing. Our results indicate that executing such strategies online involves unavoidable heat dissipation and that quantum agents can reduce this dissipation below classical limits. We found the necessary and sufficient conditions on a targeted strategy that guarantees such quantum energetic advantage and identified instances where this advantage scales without bounds.  These advantages do not require inputs or outputs to be quantum, ensuring that the quantum advantage persists when operating in purely classical environments.

A natural direction is the realization of such quantum agents. On the experimental front, recent demonstrations of Landauer's principle in quantum systems could provide a pathway to experimental validation of our results \cite{scandi2022minimally}. On the future applications front our results already give an algorithmic means to enhance the provably optimal classical counterpart thermally. A modest generalization should allow us to enhance upon existing classical agents - a key candidate being coarse-grained recurrent neural networks deployed in trade-off energy costs vs performance~\cite{zhang2019learning,he2016effective,deng2020model}. Current quantum constructions are also not necessarily optimal, indicating enhancement could be even more substantive once more optimal quantum constructions are identified \cite{liu2019optimal}. Indeed we highlight a case where increasing memory dimension can lead to improved thermodynamic performance in the Supplementary Materials. Meanwhile, our results parallel developments in agents for energy harvesting~\cite{boyd2016identifying,boyd2017leveraging,serreli2007molecular,garner2017thermodynamics}.  
  Combining these frameworks may help us tackle cases where agents harness existing temporal structure to generate more complex adaptive behavior. 
 Indeed, quantum agents - while more efficient - remain dissipative. 
 
\clearpage
\onecolumngrid
\appendix

\section{Smoothed entropies and thermodynamics}

The thermodynamics of quantum agents can be analysed via the information battery picture \cite{faist,delrio}. Before applying these techniques to quantum agents, we briefly review the definitions of quantum smoothed R\'{e}nyi min and max entropies \cite{renner2008security,faist,vitanov,delrio,tomamichel2010duality}. We start with the R\'{e}nyi max and min conditional entropies \cite{tomamichel2010duality}:

\begin{definition} The R\'{e}nyi max conditional entropy
\begin{equation}
H_{max}(B|A)_{\rho} = \max_{\omega_{A}} \log F^2(\rho_{AB}, \omega_A \otimes \openone_B)
\end{equation}
where $F(\rho_1, \rho_2) = || \rho_1^{1/2}\rho_2^{1/2}||_{tr}$ is the standard fidelity measure on quantum states (see \cite{nielsen2000quantum}). Meanwhile the R\'{e}nyi min conditional entropy is defined by
\begin{equation}
H_{min}(B|A)_{\rho} = \max_{\omega_{A}} \sup \{ \lambda \in \mathbb{R} : \rho_{AB} \le 2^{-\lambda}  (\omega_A\otimes\openone_B)\}.
\end{equation}

\end{definition}

Their smoothed counterparts are the physically relevant quantities in much of this analysis, as they can be applied to analyse the thermodynamic cost of the agent's internal map $\mathcal{T}$ under the assumption that the reset of any information batteries used in this map is implemented to within some fidelity $\varepsilon$ \footnote{Note this symbol $\varepsilon$ (varepsilon) is the smoothing parameter, and conceptually different from the encoding map $\epsilon: \past{\mathcal{Z}} \rightarrow \mathcal{S}$ in the $\epsilon$-transducer (the two concepts present with similar names and symbols as they both etymologically originate from coarse-grainings)}. The smoothed min and max entropies can be defined in terms of a purified distance, see \cite{vitanov} for details:

\begin{definition}
Let $\rho, \sigma \in S_{\le}(\mathcal{H})$, the set of subnormalised positive semi definite density operators on $\mathcal{H}$. Then  the  purified  distance between $\rho$ and $\sigma$ is defined by
\begin{equation}
P(\rho, \sigma) = \sqrt{1- F_g( \rho, \sigma)^2 }
\end{equation}
where  $F_g(\rho, \sigma) = F(\rho, \sigma) + \sqrt{(1 - \rm{tr} \, \rho)(1- \rm{tr} \,\sigma)}$
is the generalised fidelity.
\end{definition}
If either $\sigma$ or $\rho$ is a pure state we have agreement between the generalised fidelity and standard fidelity on quantum states $F_g(\rho, \sigma) = F(\rho, \sigma)$. The smoothed min and max conditional entropies can then be defined \cite{vitanov}:
\begin{definition}\label{def:smoothedentropies}
Let $\varepsilon \ge0$ and let $\rho_{AB} \in S_{\le}(\mathcal{H}_{AB})$ (the set of sub-normalised positive semi-definite density operators on the Hilbert space $\mathcal{H}_{AB}$, i.e. $\rm{tr}(\rho_{AB}) \le 1$). Then the $\varepsilon$-smooth min-entropy of $B$ conditioned on $A$ of $\rho_{AB} $ is defined as
\begin{equation}
H^{\varepsilon}_{min}(B|A)_{\rho} = \max_{\sigma} H_{min}(B|A)_{\sigma}
\end{equation}
and the $\varepsilon$-smooth max-entropy of $B$ conditioned on $A$ of $\rho_{AB}$ is defined as
\begin{equation}
H^{\varepsilon}_{max}(B|A)_{\rho} = \min_{\sigma} H_{max}(B|A)_{\sigma}
\end{equation}
where  the  maximum  and  the  minimum  range  over  all  sub-normalised states $\sigma_{AB} \approx_{\varepsilon} \rho_{AB}$ and $\rho \approx_{\varepsilon} \sigma$  iff $P(\rho, \sigma) \le \varepsilon$.
\end{definition}

The smoothed min and max entropies obey a number of useful chain rules  \cite{vitanov, tomamichel2015quantum}.
\begin{equation}
 \label{eq:ineq1}
 \begin{aligned}
&H_{max}^{\varepsilon'} (AB|C) \ge H_{max}^{\varepsilon} (A|BC) + H_{min}^{\varepsilon''}(B|C) - 3f \\
&H_{min}^{\varepsilon}(AB | C) \ge  H_{min}^{\varepsilon''}(A|BC) + H_{min}^{\varepsilon'} (B|C) - f   \\
&H_{max}^{\varepsilon} (AB|C) \le H_{max}^{\varepsilon'}(A|BC) +H_{max}^{\varepsilon''}(B|C) + f
\end{aligned}
\end{equation}
where $f \sim O(\log (1/e))$ is defined in terms of the relationship between the smoothing parameters $e = \varepsilon - \varepsilon' - 2\varepsilon ''$.

Faist et al.~showed that there is a minimum thermodynamic cost to any computational process~\cite{faist}. More specifically, if the computational process is implemented by a map $\mathcal{T}: A \rightarrow A'$, from input Hilbert space $A$ to output Hilbert space $A'$ (where $A$ and $A'$ are governed by degenerate Hamiltonians at the beginning and end of the protocol) then the thermodynamic cost of implementing $\mathcal{T}$ can be lower bounded by the following theorem:
\begin{thm}[Faist~\cite{faist}]\label{therm:Faist}
Suppose that we have a map $\mathcal{T}: A \rightarrow A'$ and that this can be realised by an isometry dilation $U : A \rightarrow A'B$ and subsequently ignoring system $B$.  Then, the minimal work cost of accomplishing this task up to an error $\varepsilon^2/2$ is at least
\begin{equation}\label{app:lowerbd}
W^{\varepsilon^2/2}/k T \ln 2 \ge  H_{max}^{\varepsilon} (B|A')
\end{equation}
assuming the Hamiltonians at the beginning and end of the protocol are degenerate.
\end{thm}
We combine this result with the following theorem from del Rio et al. \cite{delrio}: \begin{thm}[del Rio~\cite{delrio}]
There exists a process to erase a system $B$ conditioned on a memory, $A'$, acting at temperature $T$, whose work cost satisfies
\begin{equation}\label{eq:delrio}
W(B|A')\le  k T\ln 2[H^{\varepsilon}_{max}(B|A') + \Delta],
\end{equation}
except   with   probability   less   than $\delta = \sqrt{2^{-\Delta/2} + 12 \varepsilon}$  for all $\delta, \varepsilon >0$.
\end{thm}
We then obtain a constructive means of approaching the bound in Eq.~\eqref{app:lowerbd}. Namely:
\begin{theorem}
Consider a map $\mathcal{T}: A \rightarrow A'$ that can be realised by an isometry dilation $U : A \rightarrow A'B$ and subsequently ignoring system $B$. Suppose the initial and final Hamiltonians are degenerate. Then, there is a constructive mechanism for achieving this process with work cost at most
\begin{equation}\label{app:achievablebound}
W\le  k T \ln 2  \left[H_{max}^{\varepsilon} (B|A') +\Delta\right]
\end{equation}
except   with   probability   less   than $\delta = \sqrt{2^{-\Delta/2} + 12 \varepsilon}$  for all $\delta, \varepsilon >0$.
\end{theorem}

This result is described in~\cite{faist}, but is reproducible by  noting that the logical process $\mathcal{T}: A \rightarrow A'$ can be implemented by applying the unitary map  $U : A \rightarrow A'B$ (and discarding system $B$). Since $U : A \rightarrow A'B$  is unitary it is considered to be thermodynamically free when the initial and final Hamiltonians are degenerate~\cite{faist}. Thus the incurred thermodynamic cost of this protocol is entirely due to resetting the battery system  $B$ from the perspective of someone who has access to the output register $A'$. As described by del Rio et al.~\cite{delrio}, this reset operation can be done with the cost reported in Eq.~\eqref{eq:delrio} to within fidelity $\varepsilon$ with failure probability at most $\delta = \sqrt{2^{-\Delta/2} + 12 \varepsilon}$.

 In what follows we use the notation $H(A) = - \rm{Tr} \rho_A \log \rho_A$ for the Von Neumann entropy of $\rho_A$. Note that in the special case where the state $\rho_A = \sum_i P(i) \ket{i}\bra{i}$ is diagonal in the computational basis, the von Neumann entropy aligns with the Shannon entropy $H(A) =  - \sum_i P(i) \log P(i)$.  And thus we use the von Neumann entropy to analyse both the classical and quantum cases. Furthermore, we use $H(A|B) = H(AB) - H(A)$ for the quantum conditional entropy and, $I(A;B) = H(A) + H(B) - H(AB) $ for the quantum mutual information. We use the symbol $D(\rho | \sigma) =  \rm{Tr} \rho (\log \rho -\log \sigma)$ for the quantum relative entropy (for classical $\rho$ and $\sigma$ which are diagonal in the computational basis, this aligns with the classical Kullback-Leibler divergence).

\section{Quantum and classical agents}

To describe the thermodynamics of quantum agents we also need to relate details about their construction. 

In particular we assume that at each point in time $t \in \mathbb{Z}$ an agent receives an input $x_t \in \mathcal{X}$. It uses this input along with the current state of its memory $r_t \in \mathcal{R}$, to generate an output response $y_t \in \mathcal{Y}$, and update its memory to a new state $r_{t+1}$, that depends on both $z_t = (x_t, y_t)$ and $r_t$. This memory update ensures the agent remains synchronised with the history of past events, and  is now ready to repeat the above process at time step $t+1$. Here $\mathcal{X},\mathcal{Y}$ are the alphabets of admissible input, respectively output, symbols and $\mathcal{R}$ is the set of internal memory states. We require the agent's outputs to follow some desired output strategy $\mathcal{P} = \{P(Y_{0:K} = y_{0:K}| x_{0:K}, \past{z})\}_{K > 0}$  which specifies the probability with which the agent outputs $y_{0:K} = y_0y_1\ldots y_{K-1}$ when given a sequence of $K$ future inputs $x_{0:K} = x_0x_1\ldots x_{K-1}$ for each natural number $K$, conditioned on history $\past{z}$ ~\cite{barnett2015computational}.  We assume this strategy is stationary such that $P(Y_{t:t+L} | x_{t:t+L}, \past{z}_t) = P(Y_{0:L} | x_{0:L}, \past{z})$  for all $t \in \mathbb{Z}$, i.e. the distribution is time translationally invariant (however  each specific string drawn from the process, will generally not be time translationally invariant when considered in isolation.).

In both the quantum and classical case we can assume the agent's memory register starts in some well-defined distribution over memory states $\sum_i P(\sigma_i) \ket{\sigma_i}\bra{\sigma_i}$, where $P(\sigma_i)$ is the steady state distribution over memory states induced by driving of the agent system under a given i.i.d.~input sequence with entropy $h_x$.

We consider the general scenario of a $L$-stride agent that is allowed to collect and deliberate on a block of  $L$ inputs $x_{0:L} = x_0\ldots 	x_{L-1}$  before responding with a block of $L$ outputs $y_{0:L} = y_0\ldots 	y_{L-1}$ (where $L=1$ corresponds to online response). Thus the initial state of the joint tape and agent system can be described by

\begin{equation}\label{eq:initialstate}
\sum_i P(\sigma_i) \ket{\sigma_i}\bra{\sigma_i} \otimes \sum_{x_{0:L}} P(x_{0:L}) \ket{x_{0:L}}\bra{x_{0:L}} \otimes \rhodef^{\otimes L}.
\end{equation}
where $\rhodef =\frac{1}{|\mathcal{Y}|} \sum_{y \in\mathcal{Y}} \ket{y}\bra{y} $ represents the initial (maximally mixed) state of the output tape onto which the agent will transcribe its outputs. It is assumed that the state of $\rho_{\dflt}^{\otimes L}$ is governed by an i.i.d.~random variable  $\Ydef$   with entropy $L\hdef$ (where we are describing a block of $L$ units of the tape where each unit is individually i.i.d.~with entropy $\hdef$).

{\bf Classical agent} -- We consider causal agents, whose current memory state is a deterministic function of what has happened in the past $f: \past{\mathcal{Z}} \rightarrow \mathcal{R}$. When the agent is classical, we will use $R_t$ to denote the random variable governing the internal state of this agent model at time $t$. We denote the set of  internal memory states of the classical causal agent by $\mathcal{R} = \{r_i\}$. Since these internal states are orthogonal, we can always represent them in the computational basis as $r_i = |i\rangle \langle i|$.  Thus the initial state of the joint-tape and classical agent system can always described by Eq.~\eqref{eq:initialstate}, with the additional condition that $\langle \sigma_i |\sigma_j\rangle = \delta_{ij}$.

Such causal agents are also generally referred to as unifilar. This unifilarity property guarantees that if we know the internal state at time $t$, and we observe the next $L$ inputs and outputs then we know as much about the future as the agent itself, i.e. $H(R_{t+L} |  z_{t:t+L}, R_t) =0$. As a result it is possible to define a  memory update function $\lambda$ describing how the memory state updates upon observing a particular sequence of input-output pairs $z_{0:L}$.  In particular this propagator function satisfies $m' = \lambda(z_{0:L},m)$ whenever $r_m = f(\past{z})$ is the memory state corresponding to any given history $\past{z}$, and $r_m' =  f(\past{z}z_{0:L})$ that of $\past{z}z_{0:L}$.

The $\epsilon$-transducer is the classical  unifilar/causal agent that has the lowest internal entropy \cite{barnett2015computational} -- for any $\alpha$-R\'{e}nyi entropy $H_{\alpha}$, the $\epsilon$-transducer minimises the entropic quantitiy $H_{\alpha}(R_t)$, over the space of all unifilar agent models. The $\epsilon$-transducer is distinguished by an encoding function $\epsilon$ which satisfies  the relation $\epsilon(\past{z}) = \epsilon(\past{z}') $ if and only if for all possible future input strings $\future{x}$ the future output morphs of these two pasts are identical, i.e. $\epsilon(\past{z}) = \epsilon(\past{z}') $ if and only if for the strategy $\mathcal{P}$ the probability distributions $P(\future{Y} | \future{x}, \past{z}) =  P(\future{Y} | \future{x}, \past{z}')$ for all $\future{x}\in \future{\mathcal{X}}$. The internal memory states of this model are called the causal states, and generally denoted as $\mathcal{S} = \{s_i\}$. We also use  $S_t$ to denote the random variable governing the current state of the $\epsilon$-transducer, and  $P(s_i) = \pi_i$ to refer to the steady state occupation probabilities of this model's internal states \cite{barnett2015computational}. We use $P(s_j,y|x,s_i) = T^{y|x}_{ij}$ to refer to the probability an $\epsilon$-transducer initially in causal state $s_i$ emits output action $y$ and transitions to state $s_j$ upon receiving input $x$.

For any unifilar encoding functions $f$ and any two pasts $\past{z}$ and $\past{z}'$, if $f(\past{z}) = f(\past{z}')$  then  these two pasts must also satisfy the relation $\epsilon(\past{z}) = \epsilon(\past{z}')$ (i.e. if $f(\past{z}) = f(\past{z}')$ for some unifilar encoding function $f$, then $\past{z}$ and $\past{z}'$ must also be mapped to the same causal state by the $\epsilon$-transducer's encoding function). The reverse statement is not generally true~\cite{barnett2015computational}.

\begin{figure*}[tbh!]
\centering
\includegraphics[scale=0.70]{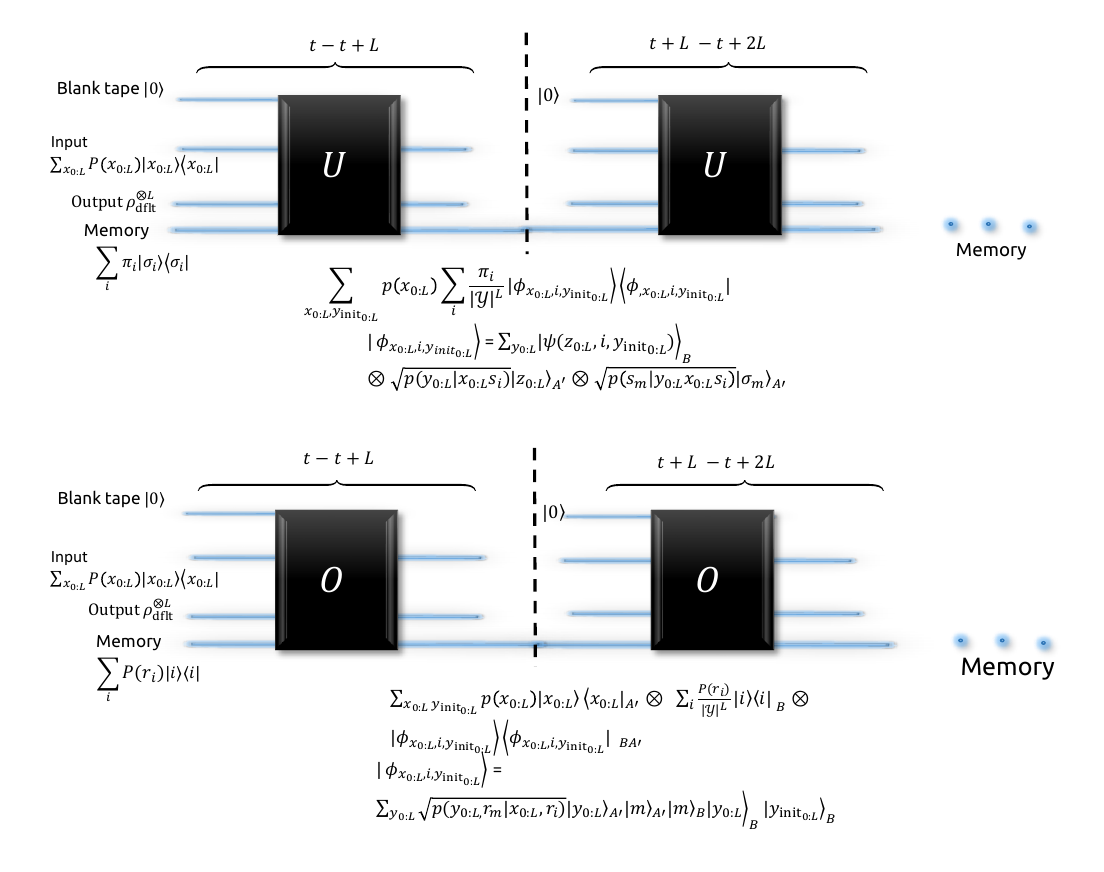}
\caption{\label{fig1} (a) A circuit indicating the dynamics of the quantum agent which at each iteration couples its internal memory register ($M_L)$ to a tape register ($Z_{0:L}$)  and a battery register (system $B$). The agent is able to thereby transduce the joint input state, of the input tape and output tape, $x_{0:L} \otimes \rhodef^{\otimes L}$ to output string $z_{0:L}$, while updating its memory to record the event. The process depletes the battery of pure states, such that at the end of each iteration the battery must be coupled to an external heat bath at some temperature $T$ and reset to its initial pure state, requiring an investment of work. (b) A classical agent implementing the same dynamics via a Stinespring dilation.}
\end{figure*}

{\bf Quantum agent -- } A quantum agent can always be implemented by a unitary map that acts jointly on the input tape system $\mathbf{X}_{0:L}$, output tape $\mathbf{Y}_{0:L}$, memory $\mathbf{M}$ and battery $\mathbf{B}$, where we set
\begin{eqnarray}\label{eq:unitaryevolution}
U\ket{x_{0:L}}_X \ket{{y_{\init_{0:L}}}}_Y\ket{\sigma_i}_M\ket{0}_B  & = &\sum_{y_{0,L},j}\sqrt{P(\sigma _j, y_{0:L} |x_{0:L}, \sigma_i)}\ket{z_{0:L}}_{Z}  \notag \\ && \ket{\sigma_{j}}_M \ket{\psi(z_{0:L},i, {y_{\init_{0:L}}})}_B
 \end{eqnarray}
provided a suitable set of junk states $\ket{\psi(z_{0:L},i, {y_{\init_{0:L}}})}_B$ can be identified, such that the resulting transformation is an isometry. Here  $\mathbf{Z}_{0:L}$ 
the input-output tape after the execution of the map, $\ket{\sigma_i}$ is the initial state of the quantum agent's memory, while $\ket{y_{{\init}_{0:L}}}$ is an arbitrary initial state of the next $L$ entries of the output tape (this output tape is assumed to be initially configured in a maximally mixed state, see Eq. \eqref{eq:initialstate}). The junk states $\ket{\psi(z_{0:L},i,  {y_{\init_{0:L}}})}_B$, represent depletion of the battery register of pure states. We can account for the thermodynamic cost of this operation, in terms of the work that must be invested by the agent to restore the battery to its initial state.

We can explicitly construct the memory states for a quantum agent via the encoding function $\epsilon_q  = \psi_q \, \circ \, \epsilon$ where $\epsilon$ is the classical encoding map from pasts onto causal states, and $\psi_q : \mathcal{S} \rightarrow \{\ket{\sigma_j}\bra{\sigma_j}\}$ replaces each classical causal state $s_j$ with a quantum counterpart $\ket{\sigma_j}\bra{\sigma_j}$.

Under these circumstances a viable set of memory states can be constructed directly from the algorithm given in~\cite{elliott2022quantum}, which associates each classical causal state $s_j$ with a quantum memory state of the form
\begin{eqnarray}\label{eq:quantumcausalstatesappendix}
 \ket{\sigma_j} &=& \otimes_{x} \ket{\sigma_j^{x}}
\end{eqnarray}
such that the overlaps $c_{ij}^x = \langle \sigma_i^{x} | \sigma_j^{x}\rangle$, are of the form
\begin{equation}\label{eq:stateoverlaps}
c^x_{ij}=\sum_y\sqrt{P(y|x,s_i)P(y|x,s_j)}\prod_{x'}c_{\lambda(z,s_i) \lambda(z, s_j)}^{x'},
\end{equation}
where $\lambda(z_{0:L} = (x_{0:L},y_{0:L}), s_i) = \epsilon(\past{x}x_{0:L}, \past{y}y_{0:L})$ for any $\past{z} = (\past{x}, \past{y} )$ such that $\epsilon(\past{x},\past{y} ) = s_i$. In particular $\lambda$ is a propagation function that computes the updated state at time $t+L$ when given the initial state at time $t$ and the  next ($L$) inputs and outputs. The correctness of this particular choice follows from \cite{Chefles2004}, along with results from \cite{elliott2022quantum} that prove there always exists a solution to the multivariate simultaneous equations \eqref{eq:stateoverlaps}.

To examine the thermodynamic cost of this implementation we set the junk states for this construction in Eq.~\eqref{eq:unitaryevolution} to
$\ket{\psi(z_{0:1}, j, {y_{\init_{0:1}}})} = \ket{y_0} \Pi_{x' \neq x_0} \ket{\sigma_j^{x'}}\ket{{y_{\init_0}}}$ for the case $L = 1$. It follows that for $L > 1$ we have
\begin{equation}
\ket{\psi(z_{0:L},j, {y_{\init_{0:L}}})} = \ket{y_{L-1}} \Pi_{x' \neq x_{L-1}} \ket{\sigma_k^{x'}} \ket{{y_{\init_{L-1}}}} \ket{\psi(z_{0:L-1}, j, {y_{\init_{0:L-1}}})}
\end{equation}
for $s_k = \lambda(z_{0:L-1},s_j)$.

Our encoding map associates each causal state $s_j$ with one quantum memory state $\ket{\sigma_j}\bra{\sigma_j}$. As a direct consequence in this quantum model we have $P(\sigma_i, y_{0:L} | x_{0:L}, \sigma_j) = P(s_i, y_{0:L} | x_{0:L}, s_j)$. Furthermore we also have alignment between the steady state causal state occupation probabilities of the causal states $\pi_i$, and the probabilities of the associated quantum memory states $P(\sigma_i) = \pi_i$  in Eq.~\eqref{eq:initialstate}.

\section{Lower bound on the work cost of executing a strategy}
To derive Result 1, we first note that by the quantum asymptotic equipartition property \cite{renner2008security} the  smoothed Renyi entropies converge to the Von Neumann entropy, in the limit of an asymptotically large ensemble of independent identically distributed (i.i.d.) copies of the state. It then follows that $H^{\varepsilon}_{max}(A|B) $ converges to $ H(A|B)$ in the i.i.d. limit. The above Theorem ({Faist} \cite{faist}) states that if  a map
$\mathcal{T}: A \rightarrow A'$ can be realised by an isometry dilation $U : A \rightarrow A'B$ and subsequently ignoring system $B$, then provided the initial and final Hamiltonians are degenerate,  in the  i.i.d. limit
\begin{eqnarray}
W&\ge&  k T \ln 2  \, H (B|A') \notag \\ &=& k T\ln 2 \left[ H(A'B) - H(A')\right]\notag \\ 
& =&  k T\ln 2 \left[ H(A) - H(A')\right] \notag \\ 
& = & (H_i - H_f) kT \ln 2 
\end{eqnarray}
where we have used $H_i = H(A) = H(U AU^{\dagger}) = H(A' B)$ for the von Neumann entropy of the input state to the map, and $H_f = H(A')$ for the entropy of the output. We note that the above relations are closely connected to Landauer's bound and similar proofs can be found in Refs. \cite{faist, riechers2021impossibility}.

For the case of our agent we recall that $H_i = L \hdef + L h_x + H(M_0)$  and $H_f = H(Z_{0:L}, M_{L})$, are the entropies of the  input and output systems respectively such that
\begin{equation}
W \ge  k T \ln 2 \left[ L h_{\textrm{dflt}} + L h_x +H(M_0) - H(Z_{0:L}, M_L)\right]
\end{equation}
Thus  observing  that
$$I(Z_{0:L}; M_{L}) - H(Y_{0:L}|X_{0:L}) = L h_x + H(M_0)  - H(Z_{0:L}, M_{L}),$$
where we have used the i.i.d.~nature of the input to set $H(X_{0:L}) = Lh_x$ and  stationarity of the construction to set $H(M_0) = H(M_L)$, we arrive at the lower bound for the energetic efficiency of any agent 
\begin{equation}
\frac{W^{(L)}}{kT \ln 2} \geq I(Z_{0:L};M_L) - H (Y_{0:L}|X_{0:L}) + L \hdef
\end{equation}
in agreement with the bound provided in Result 1.

\section{Saturating work cost bounds}\label{sec:appendixthermoagents}
 Here we outline an explicit method to saturate the bound in Result 1 for quantum agents. The construction works for general causal memory encodings, and thus applies to arbitrary quantum agents. Note that the result also applies to classical unifilar agents as a special case, and also shares elements in common with previous analysis of the work cost of quantum simulators of stochastic processes~\cite{loomis}.  Recall that before executing the policy map, the state of the joint agent-tape system is given by \eqref{eq:initialstate}. Execution of the policy map then involves applying some isometry  $U: A \rightarrow A'B$ to the joint system to yield the following output:

\begin{eqnarray}\label{eq:quantumoutput}
\rho_{A'B} &=& \sum_{x_{0:L},  y_{{\init}_{0:L}}} \sum_i P(x_{0:L})\frac{\pi_i}{|\mathcal{Y}|^L} \ket{\phi_{x_{0:L},i, {y_{\init_{0:L}}}}}\bra{\phi_{x_{0:L}, i, {y_{\init_{0:L}}}} }_{A'B} \\
\ket{\phi_{x_{0:L}, i, {y_{\init_{0:L}}}} }_{A'B} &=& \sum_{ m, y_{0:L}} \sqrt{P(y_{0:L} | x_{0:L}, \sigma_i)} \ket{z_{0:L}}_{A'}  \sqrt{P(\sigma_m|x_{0:L}, y_{0:L}, \sigma_i) } \ket{\sigma_m}_{A'} \ket{\psi(z_{0:L}, i, {y_{\init_{0:L}}})}_{B}
\end{eqnarray}

where $A'$ is the register encoding the output of the computation, and $B$ is the depleted battery which will be reset before processing the next batch of inputs.

We can thus identify the two terms $B$ and $A'$  in Eq.~\eqref{eq:quantumoutput} as being associated with marginal states 

\begin{eqnarray}
\rho_B & =& \sum_{x_{0:L} y_{{\init}_{0:L}}} \sum_i P(x_{0:L}) \frac{\pi_i}{|\mathcal{Y}|^L} \sum_{y_{0:L}} P(y_{0:L}| x_{0:L}, \sigma_i) \ket{\psi(z_{0:L},i,{y_{\init_{0:L}}} )}\bra{\psi(z_{0:L},i,{y_{\init_{0:L}}} )} \label{eq:environment}\\
\rho_{A'} & =& \sum_{x_{0:L}} \sum_i P(x_{0:L}) \pi_i \sum_{y_{0:L}} P(y_{0:L}| x_{0:L}, \sigma_i) \ket{z_{0:L}}\bra{z_{0:L}} 
\sum_{m,m'} \sqrt{P(\sigma_m| z_{0:L} ,\sigma_i) }\sqrt{P(\sigma_{m'}| z_{0:L} ,\sigma_i) }  \ket{\sigma_m}\bra{\sigma_{m'}} \notag\\
& = & \sum_{x_{0:L}} \sum_i P(x_{0:L}) \pi_i \sum_{y_{0:L}} P(y_{0:L}| x_{0:L}, \sigma_i) \ket{z_{0:L}}\bra{z_{0:L}}_{Z_{0:L}} \sum_{m} P(\sigma_m| z_{0:L} ,\sigma_i)   \ket{\sigma_m}\bra{\sigma_m}_{M_L}  \notag\\
& = & \sum_{z_{0:L}} P(z_{0:L}| \sigma_i)\pi_i \ket{z_{0:L}} \bra{z_{0:L}}_{Z_{0:L}} \otimes \sum_m P(\sigma_m | z_{0:L}, \sigma_i) \ket{\sigma_m}\bra{\sigma_m}_{M_L}.
\end{eqnarray}

Here the second-to-last line follows from unifilarity (i.e., $s_m = \lambda(z_{0:L}, s_i)$ is a deterministic function of $(z_{0:L}, s_i)$) and the relation $P(\sigma_i, y_{0:L} | x_{0:L}, \sigma_j) = P(s_i, y_{0:L} | x_{0:L}, s_j)$,  which is a direct consequence of the way we associated causal states $s_j$ in one-to-one correspondence with memory states $\ket{\sigma_j}$ in this quantum model. We also break the register $A'$ down into two sub-registers $Z_{0:L}$, and $M_L$ corresponding to the output tape and recurrent memory register. Similarly  $A$  is broken down into the input stimuli  $X_{0:L}$,  initial state of the memory register $M_0$, and initial default state of the tape the agent writes its next $L$-outputs to $\Ydef$, which has default entropy $L\hdef$.

Rearranging Eq.~\eqref{eq:ineq1} to $H_{max}^{\varepsilon}(B|A') \le  H_{max}^{\varepsilon'} (A'B)- H_{min}^{\varepsilon''}(A') + 3 f$ and substituting in Eq.~\eqref{eq:delrio} we obtain

\begin{eqnarray}
W_{q}/k T \ln 2 \, &&  \le   H_{max}^{\varepsilon} (B|A') + \Delta \\  &&  \le H_{max}^{\varepsilon'} (A'B)- H_{min}^{\varepsilon''}(A') + 3 f +\Delta. \notag
\end{eqnarray}
Noting that $A'B = UAU^{\dagger}$ for the unitary presented in Fig.~\ref{fig1}, we can thus write $H_{max}^{\varepsilon'} (A'B) = H_{max}^{\varepsilon'}(A)= H_{max}^{\varepsilon'} (X_{0:L}, M_0, \Ydef)$. Furthermore we can use Eq.~\eqref{eq:ineq1} with the second term and make suitable choices for $\varepsilon', \varepsilon''$ etc to obtain

\begin{eqnarray}
W^{\varepsilon}_q/k T \ln 2 & \le & H_{max}^{\varepsilon/4} (X_{0:L}, M_{0}, \Ydef)- H_{min}^{\varepsilon/4}(M_L,Z_{0:L})  + O(\log(1/\varepsilon)) +\Delta \label{eq:quantumoneshotcost} \\
&=& H_{max}^{\varepsilon/4} (X_{0:L}, M_{0}, \Ydef)- (H_{min}^{\varepsilon/16}(M_L|Z_{0:L}) + H_{min}^{\varepsilon/16} (Z_{0:L})) + O(\log(1/\varepsilon)) +\Delta \notag \\
& = & H_{max}^{\varepsilon/4} (X_{0:L}) + H_{max}^{\varepsilon/4}( M_{0}) + H_{max}^{\varepsilon/4}( \Ydef)- H_{min}^{\varepsilon/16}(M_L|Z_{0:L}) - H_{min}^{\varepsilon/16} (Z_{0:L}) + O(\log(1/\varepsilon)) +\Delta, \notag
\end{eqnarray}

where we have used the fact that the R\'{e}nyi max entropy is additive when the two systems are uncorrelated.

Finally we take the i.i.d.~limit where we process many copies in parallel, such that we can operate on $\rho_{A'B}^{\otimes n}$ in the limit of large $n$. Here the smoothed conditional min and max entropies converge to the von Neumann entropy such that the work cost of a single agent emitting $L$ outputs in the i.i.d.~limit can be simplified to
\begin{eqnarray}
W^{(L)}_q/k T \ln 2 &\le& H(X_{0:L}) + H(M_{0}) + L\hdef \label{eq:iidquantumcostineq} \\ &&  - H(M_L|Z_{0:L}) -H(X_{0:L}, Y_{0:L}) \notag \\
 &=&  L\hdef - H(Y_{0:L} | X_{0:L}) + I(Z_{0:L}; M_L). \notag
 \end{eqnarray}
In the last line we have assumed the memory starts and ends in state $\sum_i \pi_i \ket{\sigma_i}\bra{\sigma_i}$ -- this amounts to the stationarity assumption -- i.e. $P(Y_{t:t+L} | x_{t:t+L}, \past{z}_t) = P(Y_{0:L} | x_{0:L}, \past{z})$  for all $t \in \mathbb{Z}$ -- such that there is no sudden discontinuity in the distribution over driving  input sequences or conditional output response behaviour, at time $t = 0$.
Meanwhile the terms $\Delta$ and $O(\log(1/\varepsilon))$ can be made arbitrarily small in the i.i.d.~scenario~\cite{delrio}.

We see that this coincides with the lower bound for the i.i.d.~work cost from Result 1 for executing this strategy. Therefore, the upper and lower bounds coincide, and thus we can set the inequality in equation \eqref{eq:iidquantumcostineq} to an equality, and obtain
\begin{equation}
W^{(L)}_q/k T \ln 2 =  L\hdef - H(Y_{0:L}|X_{0:L})  + I(Z_{0:L}; M_L). \label{eq:iidquantumcost}
\end{equation}
as both the necessary and achievable i.i.d. work cost of a quantum agent producing $L$ output responses.

\section{Thermal optimality of $\epsilon$-transducers}
Here, we establish that $\epsilon$-transducers have minimal work cost among all classical agents. Recall the thermodynamics of classical agents can also be derived using the information battery paradigm (see Fig.~\ref{fig1}(b)). This allows any classical unifilar agent with encoding function $f: \past{\mathcal{Z}} \rightarrow \mathcal{R}$, to be directly analysed as a special case of the quantum construction. We use the symbol $W_r^{(L)}$ to denote the i.i.d.~work cost of producing $L$-outputs with an agent with encoding function $f:\past{\mathcal{Z}} \rightarrow \mathcal{R} = \{r_i\}$ (assuming the agent can generate $L$-outputs at a time):
\begin{equation}
W^{(L)}_r/k_B T \ln 2  = L\hdef - H(Y_{0:L}| X_{0:L})  + I(Z_{0:L}; R_L), \label{eq:iidclassicalcostnonunifilar}
 \end{equation}
provided the  agent starts and ends in the same steady state  distribution over memory states (which is a natural consequence of the i.i.d.~driving process $\omni{X}$ and stationarity of the strategy $\mathcal{P}$). Here  we have assigned labels $R_L$ and $Z_{0:L}$ to the random variables governing the memory and output registers respectively, and $\hdef$ is the per symbol entropy of the initial state of the output tape to which the agent transcribes its output responses $Y_t$.

There is a compressive map that relates the states of any unifilar classical agent $\mathcal{R} = \{r_i\}$ to the state of the $\epsilon$-transducer $\mathcal{S} = \{s_i\}$. In particular if two pasts $\past{z},\past{z}'$, are mapped to the same internal memory state of the unifilar model $f(\past{z}) = f(\past{z}')$, then both pasts will also belong to same causal state $\epsilon(\past{z}) = \epsilon(\past{z}')$   \cite{barnett2015computational}. This implies there is a Markov chain relation mapping $\past{z} = (\past{x}, \past{y}) \rightarrow r_j \rightarrow s_i$.  We will use this fact and the data processing inequality to show
$I(Z_{-L:0}; R_0)  \ge I(Z_{-L:0}; S_0) $. Due to stationarity, this implies $I(Z_{0:L}; R_L)  \ge I(Z_{0:L}; S_L)$  where $S_L$ is the random variable governing the causal state at time $t= L$. To do this, we make use of 
 
\begin{itemize}
\item The chained conditional mutual information relations $I(A_1 A_2;B|C) = I(A_1;B|C) + I(A_2;B|CA_1)$.
\item The relation  $I(A_2 ; B |A_1) = 0$ whenever there is a physical channel $g_{A_1 \rightarrow A_1A_2}$ such that $\rho_{A_1 A_2 B} = g_{A_1\rightarrow A_1A_2} (\rho_{A_1 B})$ \cite{fawzi2015quantum,renner2008security}.
\end{itemize}

Here $I(A_1;B|C) = H(A_1, C)  + H(B,C) - H(C) - H(A_1 ,B ,C)$ and $H(A_1|B) = H(A_1 B) - H(B)$  \cite{fawzi2015quantum}. In particular we observe that from the map $\past{z} = (\past{x}, \past{y}) \rightarrow   r_j$ we inherit a well-defined joint state $ \sum_{\past{z}}P(r_j|\past{z}) P(\past{z}) r_j \otimes \ket{\past{z}}\bra{\past{z}}$, tracing out all time steps before time $t = -L$ and assigning labels $A_1, B$ to the memory and tape subspaces respectively yields
\begin{eqnarray}\label{eq:inputMarkovchain}
\rho_{A_1 B} = \sum_{z_{-L:0}} \left(\sum_{r_j}P(r_j|z_{-L:0}) \, {r_j}\right)_{A_1}  \otimes P(z_{-L:0}) \ket{z_{-L:0}}\bra{z_{-L:0}}_B.   
\end{eqnarray}
Due to the existence of a Markov chain mapping  $\past{z} = (\past{x}, \past{y}) \rightarrow   r_j \rightarrow s_i $, we can build a channel  $g_{A_1 \rightarrow A_1A_2}$ which maps the state in Eq.~\eqref{eq:inputMarkovchain} to
\begin{eqnarray}\label{eq:mutualinformationjointstate}
\rho_{A_1 A_2 B} = \sum_{z_{-L:0}}\left( \sum_{r_j} P(r_j|z_{-L:0}) \, r_j \otimes \sum_{s_i} P(s_i|r_j) \, s_i\right)_{A_1A_2}\otimes   P(z_{-L:0}) \ket{z_{-L:0}}\bra{z_{-L:0}}_B    
\end{eqnarray}
Thus we can apply our chained conditional mutual information inequalities plus the data processing inequality to write  $I(R_0 S_0; Z_{-L:0}) = I(R_0; Z_{-L:0}) + I(S_0; Z_{-L:0}|R_0) = I(R_0; Z_{-L:0})$. It directly follows from this observation that $I(R_0; Z_{-L:0}) =I(R_0 S_0; Z_{-L:0})  \ge I(S_0; Z_{-L:0})$. Thus we have established that $I(R_L; Z_{0:L}) \ge I(S_L; Z_{0:L})$.

This implies that in the i.i.d.~limit we find the work cost follows a hierarchy 
  \begin{equation}
  W^{(L)}_r  \ge  W^{(L)}_{c} 
 \end{equation}
 where $W^{(L)}_c$ is the work cost of using the classical $\epsilon-$transducer agent model to execute this task with an $L$-stride window (i.e. the cost of producing $L$ symbols using the $\epsilon-$transducer, when the agent is allowed to collect $L$-inputs and emit all $L$ corresponding-output-responses at the same time). This is true for every finite $L$.

 This result has two consequences. First it places an ultimate limit on the efficiency of any classical unifilar construction operating in the i.i.d.~regime
 \begin{equation}
W^{(L)}_c/k_B T \ln 2  = L\hdef - H(Y_{0:L}| X_{0:L})  + I(Z_{0:L}; S_L).
 \end{equation}
It simultaneously implies this limit can be saturated by using the $\epsilon$-transducer agent construction, and information battery protocol introduced in the quantum agents section \cite{faist, delrio}.

\section{Proof of thermodynamic advantage for Quantum Agents}

Here, we establish the conditions in which quantum agents are guaranteed to have an energetic advantage, culminating in Result 2. We begin by showing the techniques used at the end of the last section directly imply:
\begin{equation}
W^{(L)}_{c} \ge W^{(L)}_q.
\end{equation}
That is, for any fixed $L$, the minimum thermodynamic cost of generating responses with any  $L$-stride classical agent $W_c^{(L)}$, always upper bounds the cost of its quantum counterpart $W_q^{(L)}$.

We directly obtain this result from the structure of the quantum encoding function $\epsilon_q = \psi_q \circ \epsilon$. That is $\epsilon_q$ involves a compressive map from the causal states onto the quantum memory states, i.e. $\psi_q: s_i \rightarrow \ket{\sigma_i}\bra{\sigma_i}$. As a result it is always possible to refactor the encoding maps from pasts onto memory states of the quantum agent according to a series of deterministic functions  $\past{z}=(\past{x}, \past{y}) \rightarrow   r_j \rightarrow s_i \rightarrow |\sigma_i\rangle\langle \sigma_i| $ \cite{barnett2015computational, thompson2017using, elliott2022quantum}. Therefore the above logic used to show $W^{(L)}_r  \ge  W^{(L)}_{c}$ holds. In particular, we can always recover the state $\sum_{z_{-L:0}} P(s_i|z_{-L:0}) \, s_i\otimes  \ket{z_{-L:0}}\bra{z_{-L:0}}$ as a marginal of \eqref{eq:mutualinformationjointstate}, and use the Markov chain $\psi_q: s_i \rightarrow |\sigma_i\rangle\langle \sigma_i| $ to show $I(S_0 M_0; Z_{-L:0}) = I(S_0; Z_{-L:0}) + I(M_0; Z_{-L:0}|S_L) = I(S_0; Z_{-L:0})$. It follows directly from this observation that $I(S_0; Z_{-L:0}) = I(S_0 M_0; Z_{-L:0}) \ge I(M_0; Z_{-L:0})$.  

Next, our goal is to show that the gap
\begin{equation}\label{eq:thermodynamicadvantage}
 w^{(L)}_c - w^{(L)}_q = k T \ln 2/L\left[I(Z_{0:L}; S_L) - I(Z_{0:L}; M_L)\right],
\end{equation}
is strictly positive if and only if the agent has a causally wasteful pair of memory states $s_j, s_k$ and $P(s_j|z_{0:L}) \neq P(s_j)$ for some past $z_{0:L}$. To do this we  need to make use of two different results. 

First we need to formalize the concept of a causally wasteful pair $s_j, s_k$. Using the framework of \cite{thompson2017using}, we adopt the format of a two player game, in which an interrogator Bob, is asking Alice the agent (namely the $\epsilon$-transducer) questions. At time $t = 0$, Bob is promised Alice's memory is either in state $s_j$ or $s_k$. Recall that we say $s_j$ and $s_k$ is a causally wasteful pair if there is no way for Bob to know with certainty whether Alice's memory started in state $s_j$ or $s_k$ via any means of interrogating Alice (i.e., asking her questions).

To specify this formally, we first introduce a mathematical definition for an \emph{interrogation strategies}. Let the interrogation begin at $t = 0$. At each time $t \geq 0$, Bob can ask Alice an input question $x_t$ of his choice, resulting in corresponding responses $y_t$ whose statistics are governed by Alice's $\epsilon$-transducer. Bob can base his decision for each $x_t$ on (1) all past inputs $x_{0:t}$, (2) all past outputs $y_{0:t}$ and (3) explicit time-dependence $t$. Unlike Alice, Bob does not have memory constraints and can thus execute non-stationary strategies. An interrogation strategy then defines the most general action sequence Bob can take:

\begin{definition}[Interrogation Strategy]
    An interrogation strategy $\Lambda$ is then a family of probability distributions $\{\Lambda_t(X_t = x_t | z_{0:t}), t = 0,1,\ldots\}$, specifying the probability Bob will decide on input question $x_t$ upon seeing past history $z_{0:t} = (x_{0:t}, y_{0:t})$. 
\end{definition}

Suppose Alice is initially deployed in state $s_j$, each interrogation strategy $\Lambda$ then results in a sequence of output responses $y_0,y_1,\ldots$, governed by the conditional probability distribution $P_{\Lambda}(\future{Y}|s_j)$. Bob is then unable to determine with certainty where Alice was initially in state $s_j$ or $s_k$ provided $\sum_{\future{y}} P_\Lambda (\future{y} | \future{x},s_j) P_\Lambda(\future{y} |\future{x},s_k) > 0$. Thus we say that $s_j, s_k$ is a causally wasteful pair if and only if  for all interrogation strategies $\Lambda$ we have $\sum_{\future{y}} P_\Lambda(\future{y} | \future{x},s_j) P_\Lambda(\future{y} |\future{x},s_k) > 0$. This implies there is no strategy $\Lambda$ which Bob can use to decide whether Alice was initially in $s_j$ or $s_k$ and win the game with certainty.

The second result we need to invoke is the Petz recovery map \cite{petz1988sufficiency,petz1986sufficient, ruskai2002inequalities}. In particular we make use of the following statements about the Petz recovery map and monotonicity of the data processing inequality 

\begin{thm}[Ruskai ~\cite{ruskai2002inequalities}]\label{thm:Ruskai}
Consider  monotonicity of the relative entropy $D(\rho | \sigma) \ge  D(\Phi(\rho)|\Phi(\sigma))$ where $\Phi$ is a CPTP map, and $D(\rho|\sigma)$ is the quantum relative entropy. Equality  $D(\rho | \sigma) = D(\Phi(\rho)|\Phi(\sigma))$ holds if and only if 

\begin{equation}\label{eq:Ruskai1}
\log \rho - \log \sigma = \hat{\Phi} \left[\log \Phi(\rho) - \log\Phi(\sigma)\right],
\end{equation}
where $\hat{\Phi}$  is the adjoint map to $\Phi$ and is defined by $\rm{Tr}(A^{\dagger} \hat{\Phi}(B)) = \rm{Tr}(\Phi(A)^{\dagger} B)$.

Furthermore a necessary condition for equality $D(\rho | \sigma) = D(\Phi(\rho)|\Phi(\sigma))$ is 
\begin{equation}\label{eq:Ruskai2}
\Phi(\log \rho - \log \sigma ) = \Phi(I) \left[\log \Phi(\rho) - \log\Phi(\sigma)\right],
\end{equation}
where $I$ is the Identity matrix.
\end{thm}

\textbf{Proof of Result 2} -- We will start by proving the forward direction of our if and only if statement in Result 2. i.e., we prove that if there exists a causally wasteful pair $s_i, s_k$ and $P(s_i|z_{0:L}) \neq P(s_i)$ for some $z_{0:L}$, then $w^{(L)}_c - w_q^{(L)} >0$. i.e., there exist some quantum agent which has a non-zero thermodynamic advantage over all classical agents (in particular the quantum agents in \cite{elliott2022quantum} and \cite{thompson2017using} will both have such an advantage)

 To do this we use proof via the contrapositive. That is we prove that if $w^{(L)}_c - w_q^{(L)} = 0$ then for all $s_i\in\mathcal{S}$ either (a) there cannot exist any causally wasteful pairs $s_i, s_k$, or (b)  $P(s_i) = P(s_i|z_{0:L})$ for all $z_{0:L}$.

 We thus begin by assuming $w^{(L)}_c - w_q^{(L)} = 0$. We can rewrite this condition in terms of the entropies of the memory distributions conditioned on seeing the last $L$ symbols $z_{0:L}$. For the classical agent this conditional memory state is $\rho_{c|z_{0:L}} = \sum_{s_i}  P(s_i|z_{0:L})\ket{i}\bra{i}$, and for its quantum counterpart the conditional memory state is $\rho_{q|z_{0:L}} = \sum_{s_i}  P(s_i|z_{0:L})\ket{\sigma_i}\bra{\sigma_i}$, such that

\begin{eqnarray}\label{eq:quantumadvantage}
0=  (w^{(L)}_c -w^{(L)}_q)L/(k T\ln 2) & = & I(S_L; Z_{0:L}) - I(M_L; Z_{0:L})  \notag \\
&=&  H(S_L ) - H(M_L)  - H(S_L|Z_{0:L}) + H(M_L|Z_{0:L}) \notag \\
& =& \sum_{z_{0:L}} P(z_{0;L}) D(\rho_{c|z_{0:L}} | \rho_c) - \sum_{z_{0:L}} P(z_{0:L}) D(\rho_{q|z_{0:L}}| \rho_q)\notag \\
& =& \sum_{z_{0:L}} P(z_{0;L}) \left[D(\rho_{c|z_{0:L}} | \rho_c) -  D({\psi_q}(\rho_{c|z_{0:L}})|{\psi_q}( \rho_c)) \right]  \end{eqnarray}
where $D(\rho|\sigma) = \textrm{Tr}\rho[ \log \rho - \log \sigma]$ is the quantum relative entropy and  $\psi_q(\rho) = \sum_k F_k \rho F_k^{\dagger}$ for $F_k = |\sigma_k\rangle \langle k|$, is the channel that maps each causal state $s_i = \ket{i}\bra{i}$ to its quantum counterpart $\ket{\sigma_i}\bra{\sigma_i}$. We have also labeled the mixed state $\rho_{c} = \sum_{z_{0:L}}P(z_{0:L}) \rho_{c|z_{0:L}}$  (with a similar label for $\rho_q = \sum_{z_{0:L}}P(z_{0:L}) \rho_{q|z_{0:L}}$). Note that due to monotonicity of the relative entropy we immediately have that for each $z_{0:L}$,

\begin{equation}\label{eq:dataprocessinginequality}
D(\rho_{c|z_{0:L}} | \rho_c) -  D({\psi_q}(\rho_{c|z_{0:L}})|{\psi_q}( \rho_c)) \ge 0.
\end{equation}
It follows immediate that $w_{c}^{(L)} - w_q^{(L)} = 0 $ if and only if  $ D(\rho_{c|z_{0:L}} | \rho_c) -  D({\psi_q}(\rho_{c|z_{0:L}})|{\psi_q}( \rho_c)) = 0$ for all $z_{0:L}$.

This means the data processing inequality is independently saturated  for each $z_{0:L}$. Implying that for each $z_{0:L}$ we have one equation of the form given in Eq.  \eqref{eq:Ruskai1}. These conditions correspond to setting $\rho = \rho_{c|z_{0:L}}$, $\sigma = \rho_c$ and $\Phi = \psi_q = \sum F_k \cdot F_k^{\dagger}$ for $F_k = \ket{\sigma_k}\bra{k}$ and $\ket{\sigma_k}$ defined in Eq. \eqref{eq:quantumcausalstatesappendix}, such that $\Phi(\rho) = \rho_{q|z_{0:L}}$ and $\Phi(\sigma) = \rho_q$ in Eq. \eqref{eq:Ruskai1}.  This translates into (for all $z_{0:L}$)

\begin{equation} \label{eq:rusakiimplication1}
\sum_i \left(\log P(s_i|z_{0:L}) - \log P(s_i) \right) \ket{i}\bra{i}= \hat{\Phi} \left[\log \sum P(s_i|z_{0:L}) \ket{\sigma_i} \bra{\sigma_i} - \log\sum P(s_i) \ket{\sigma_i} \bra{\sigma_i} \right].
\end{equation}
This a matrix equation and must be satisfied element-wise. This implies that for each $j$, $k$, the condition
\begin{eqnarray}
\rm{Tr} \left(\ket{k}\bra{j} \sum_i \left(\log P(s_i|z_{0:L}) - \log P(s_i) \right) \ket{i}\bra{i}\right) &= & \left(\log P(s_k|z_{0:L}) - \log P(s_k) \right) \delta_{kj}, \notag\\
&=& \rm{Tr} \left(\Phi(\ket{j}\bra{k})^{\dagger} \left[\log \sum P(s_i|z_{0:L}) \ket{\sigma_i} \bra{\sigma_i} - \log\sum P(s_i) \ket{\sigma_i} \bra{\sigma_i} \right]\right), \notag \\
& = & \langle \sigma_j | \left[\log \rho_{q|z_{0:L}} - \log\rho_{q}  \right] |\sigma_k\rangle. \label{eqn:mateq}
\end{eqnarray}
We simultaneously find  Eq. \eqref{eq:Ruskai2} implies that for each $z_{0:L}$ we must also have
\begin{eqnarray}
\sum \left(\log P(s_i|z_{0:L}) - \log P(s_i) \right) \ket{\sigma_i}\bra{\sigma_i}=  \left[\sum_{k} \ket{\sigma_k}\bra{\sigma_k}\right]  \left[\log \sum P(s_i|z_{0:L}) \ket{\sigma_i} \bra{\sigma_i} - \log\sum P(s_i) \ket{\sigma_i} \bra{\sigma_i} \right].
\end{eqnarray}
We now take $\rm{Tr} \left[ \ket{\sigma_m}\bra{\sigma_m} \cdot \right]$ on both the left and right hand side of the above equation. On the left hand side, we have
\begin{eqnarray}
\rm{Tr} \left[ \ket{\sigma_m}\bra{\sigma_m} \sum_i \left(\log P(s_i|z_{0:L}) - \log P(s_i) \right) \ket{\sigma_i}\bra{\sigma_i} )\right]& =&  \sum_i \left(\log P(s_i|z_{0:L}) - \log P(s_i) \right) |\langle \sigma_i |\sigma_m\rangle|^2  \label{eqn:LHS}
\end{eqnarray}
Meanwhile, on the right hand side, we have
\begin{eqnarray}
\rm{Tr} \left[ \ket{\sigma_m}\bra{\sigma_m} \left[\sum_{k} \ket{\sigma_k}\bra{\sigma_k}\right] \times \left[\log \rho_{q|z_{0:L}} - \log\rho_{q}  \right]\right]\notag  
& =& \sum_{k} \langle \sigma_m |\sigma_k \rangle \bra{\sigma_k}\left[\log \rho_{q|z_{0:L}} - \log\rho_{q}  \right]\ket{\sigma_m} \notag \\ 
&= & \sum_{k} \langle \sigma_m |\sigma_k \rangle  \left(\log P(s_k|z_{0:L}) - \log P(s_k) \right) \delta_{km} \notag \\ 
& = &   \log P(s_m|z_{0:L}) - \log P(s_m),
\label{eqn:RHS}
\end{eqnarray}
where we have made use of the first and last line of Eq. (\ref{eqn:mateq}). Since Eqns. (\ref{eqn:LHS}) and (\ref{eqn:RHS}) are equal, and the last line of (\ref{eqn:RHS}) represents a single term of the sum in (\ref{eqn:LHS}), the remainder of the sum must be zero. Thus, for each $z_{0:L}$
\begin{equation}\label{eq:conditionsfornonrecovery}
\sum_{i \neq m} \left(\log P(s_i|z_{0:L}) - \log P(s_i) \right) |\langle \sigma_i |\sigma_m\rangle|^2  = 0.
\end{equation}
We can then take a convex combination of these conditions in Eq. \eqref{eq:conditionsfornonrecovery}, weighted by the coefficients $P(z_{0:L})$ to get
\begin{equation}\label{eq:overlapconditions}
\sum_{i \neq m} \left(\sum_{z_{0:L}} P(z_{0:L})\log P(s_i|z_{0:L}) - \log \left( \sum_{z_{0:L}} P(z_{0:L}) P(s_i|z_{0:L}) \right) \right) |\langle \sigma_i |\sigma_m\rangle|^2  = 0.
\end{equation}
But by concavity of the logarithm we have $\sum_{z_{0:L}} P(z_{0:L})\log P(s_i|z_{0:L}) - \log \left(\sum_{z_{0:L}} P(z_{0:L}) P(s_i|z_{0:L}) \right)\ge 0$ for all $i$. Equality in Eq. \eqref{eq:overlapconditions} would thus imply each term in the sum over $i$ is independently equal to zero. That is we must have 
\begin{equation}
    \left(\sum_{z_{0:L}} P(z_{0:L})\log P(s_i|z_{0:L}) - \log \left(\sum_{z_{0:L}} P(z_{0:L}) P(s_i|z_{0:L}) \right) \right) |\langle \sigma_i |\sigma_m\rangle|^2  = 0
\end{equation}
for each $m$, and every $i\neq m$. It follows that for all $\ket{\sigma_i}$ we must have

\begin{itemize}
    \item$|\langle \sigma_i |\sigma_m\rangle|^2 = 0$ for all $m \neq i$ or,
    \item $\sum_{z_{0:L}} P(z_{0:L})\log P(s_i|z_{0:L}) - \log P(s_i) = 0$.
\end{itemize} 
Thus for all causal states $s_i\in\mathcal{S}$ either (i) the corresponding quantum memory state $\ket{\sigma_i}$ is orthogonal to all other memory states or (ii) $P(s_i|z_{0:L}) = P(s_i)$ for all $z_{0:L}$ such that the last $L$ inputs and outputs contain no information about whether the agent is in state $s_i$. 

To finish the proof we simply need to show that there exists a quantum model with $|\langle \sigma_i | \sigma_m\rangle |^2 > 0 $, if and only if $s_i, s_m$ are \emph{a  causally wasteful pair}.

To do this we adopt a result from \cite{thompson2017using} Theorem 1, which shows that  $|\langle \sigma_i | \sigma_m \rangle |^2 = 0 $ for all quantum models, if and only if there exists an interrogation strategy $\Lambda$ such that $D(P_{\Lambda}(\future{y} |s_i), P_{\Lambda}(\future{y} |s_m))  = 1$ where $D(\cdot, \cdot)$ is the trace distance -- note that since  $1-F(\rho, \sigma) \le D(\rho, \sigma) \le \sqrt{1-F(\rho,\sigma)^2}$ \cite{nielsen2000quantum} the condition $D(P_{\Lambda}(\future{y} |s_i), P_{\Lambda}(\future{y} |s_m))  = 1$ is equivalent to $F(P_{\Lambda}(\future{y} |s_i), P_{\Lambda}(\future{y} |s_m))  = 0$.   We can thus immediately rephrase the results of \cite{thompson2017using}  as $|\langle \sigma_i | \sigma_m \rangle |^2 = 0 $ for all quantum models, if and only if there exists $\Lambda$ such that $\sum_{\future{y}} P_\Lambda(\future{y} | \future{x},s_m) P_\Lambda(\future{y} |\future{x},s_i) = 0$. That is  $|\langle \sigma_i | \sigma_m \rangle |^2 = 0 $ for all quantum models, if and ony if  $s_i, s_m$  is \emph{not a causally wasteful pair.}

 Putting these statements together we arrive at the implication if $w_{c}^{(L)} - w_q^{(L)} = 0 $ for all quantum agents, then for all $s_i\in\mathcal{S}$ either (a
) there can not exist any causally wasteful pairs $s_i, s_m$ or (b) $P(s_i) = P(s_i |z_{0:L})$ for all $z_{0:L}$.
 
\textbf{Proof of reverse direction in Result 2}  -- 
Finally we establish the reverse direction of our if and only if statement. That is we prove that if  $w^{(L)}_c - w_q^{(L)} >0$, then there exists a causally wasteful pair $s_i, s_k$ and $P(s_i|z_{0:L}) \neq P(s_i)$ for some $z_{0:L}$. We again use the method of proof by the contrapositive. That is we prove that if for all $s_i\in \mathcal{S}$  either (a) there exists no causally wasteful pairs $s_i, s_m$ or (b) $P(s_i) = P(s_i |z_{0:L})$ for all  $z_{0:L}$, then $w_c^{(L)} - w_q^{(L)} = 0 $ for all quantum agents. To do this we start by assuming that for all $s_i\in \mathcal{S}$  either (a) there exists no causally wasteful pairs $s_i, s_m$ or (b) $P(s_i) = P(s_i |z_{0:L})$ for all  $z_{0:L}$. This motivates the construction of two sets $A, B \subset \mathcal{S}$ of memory states. Set $A$ contains all $s_i$ such that there are no causally wasteful pairs of the form $s_i, s_m$. Set $B$ is defined by $B =\{s_i \,\, |  \,\, P(s_i) = P(s_i | z_{0:L}) \textrm{ for al } z_{0:L} \textrm{  and  } s_i \notin A\}$. By the assumptions above we have 
\begin{equation}\label{eqn:setlaws}
\mathcal{S} = A \cup B \qquad A \cap B = \emptyset. 
\end{equation}
Any classical agent would encode each causal $s_k$ into a mutually orthogonal quantum state $\ket{k}$, which we define as the computational basis. Consider now any quantum agent that executes the equivalent strategy, where each causal state $s_k$ is encoded within some corresponding quantum memory state $\ket{\sigma_k}\bra{\sigma_k} = \psi_q(\ket{k}\bra{k})$, such that $\psi_q(\rho) = \sum_k F_k \rho F_k^{\dagger}$ for $F_k = |\sigma_k\rangle \langle k|$. By Theorem 1 of  \cite{thompson2017using}
$$A = \{ s_i \,\,|  \,\langle \sigma_i | \sigma_m\rangle = \delta_{im} \textrm{ for all } s_m\in\mathcal{S} \}.$$ and in particular, the quantum memory states of all causal states in $A$ must be mutually orthogonal. Therefore we can represent each quantum memory state $\ket{\sigma_i}$ in $A$ by a corresponding computational basis state $\ket{i}$. That is $\psi_q$ cannot compress states in $A$.

Let $\hat{\Phi}$ be the adjoint of $\psi_q$ such that $\hat{\Phi}(\rho) = \sum_k F_k^\dagger \rho F_k$ . In the the proof in the forward direction, we established that
\begin{equation}\label{eqn:reverse}
\log\rho_{c|z_{0:L}} - \log \rho_c = \hat{\Phi} \left[\log \rho_{q|z_{0:L}} - \log \rho_q\right]. \end{equation}
for all $z_{0:L}$ if and only if $w_c^{(L)} = w_q^{(L)}$. So if we can prove Eq.  \eqref{eqn:reverse} holds then this directly establishes $w_c^{(L)} = w_q^{(L)}$. These equations are in terms of $\rho_{c|z_{0:L}} = \sum_{s_i}  P(s_i|z_{0:L})\ket{i}\bra{i}$ and $\rho_c = \sum_{z_{0:L}} P(z_{0:L}) \rho_{c|z_{0:L}}$, and their quantum counterparts $\rho_{q|z_{0:L}} = \sum_{s_i}  P(s_i|z_{0:L})\ket{\sigma_i}\bra{\sigma_i}$ and $\rho_q = \sum_{z_{0:L}} P(z_{0:L}) \rho_{q|z_{0:L}}$.

Eqn. (\ref{eqn:setlaws}) allows us to rewrite the terms in Eq. \eqref{eqn:reverse} as $\rho_c = \rho_{c,A} \oplus \rho_{c,B}$, and  $\rho_{c|z_{0:L}} = \rho_{c|z_{0:L}, A} \oplus \rho_{c|z_{0:L}, B}$ where by $\rho_{r,A}$, we mean $\rho_r$ projected onto the subspace spanned by causal states in $A$, and $\rho_{r,B}$ is defined analogously. 

The above direct sum structure is respected under $\psi_q$ such that $\rho_q  = \psi_q(\rho_{c}) =\psi_q(\rho_{c,A}) \oplus \psi_q(\rho_{c,B})  = \rho_{q,A} \oplus \rho_{q,B}$ and  $\rho_{q|z_{0:L}} = \psi_q(\rho_{c|z_{0:L}}) = \psi_q(\rho_{c|z_{0:L}, A}) \oplus \psi_q(\rho_{c|z_{0:L}, B}) = \rho_{q|z_{0:L}, A} \oplus \rho_{q|z_{0:L}, B }$. Note that by the definition of $A$ we will automatically have $\rho_{c,A} = \rho_{q,A}$ and $\rho_{c|z_{0:L}, A} = \rho_{q|z_{0:L}, A} $ for all $z_{0:L}$. Meanwhile due to the definition of $B$ for all $z_{0:L}$, we also automatically have $\rho_{c,B}  = \rho_{c|z_{0:L}, B}$ and $\rho_{q,B}  = \rho_{q|z_{0:L}, B}$. 

Since this Eq. (\ref{eqn:reverse}) is a matrix equation we must have equality on an entry by entry basis. This means $w_c^{(L)} - w_q^{(L)} = 0 $ if and only if for all $z_{0:L}$ and every pair $s_i, s_k$ we have
\begin{equation}\label{eq:breakdownofrusaki1}
\langle i | \log\rho_{c|z_{0:L}, A} \oplus \log\rho_{c|z_{0:L}, B}  - \log \rho_{c,A} \oplus \log \rho_{c,B} |k\rangle  
= \langle \sigma_k | \log \rho_{q|z_{0:L}, A} \oplus \log \rho_{q|z_{0:L}, B} - \log  \rho_{q,  A} \oplus \log\rho_{q, B} |\sigma_i\rangle
\end{equation}

We now show that \eqref{eq:breakdownofrusaki1}  is true. To do this we break this set of conditions down into three cases. Case (1) we have $s_i, s_k \in A$. Case (2) we have $s_i \in A$ and $s_k \in B$. Case (3) $s_i, s_k \in B$.

We consider Case (1) first. Here both $s_i, s_k \in A$ and we can reduce the above equation to 
\begin{eqnarray}
\langle i | \log\rho_{c|z_{0:L}, A} \oplus \log\rho_{c|z_{0:L}, B}  - \log \rho_{c,A} \oplus \log \rho_{c,B} |k\rangle   & = & (\log P(s_i |z_{0:L}) - \log P(s_i)) \delta_{ik} \\ &=&  \langle k | \log \rho_{c|z_{0:L}, A}  - \log  \rho_{c,  A} |i \rangle \notag \\
&=& \langle k | \log \rho_{q|z_{0:L}, A}  - \log  \rho_{q,  A} |i \rangle \notag \\
&=& \langle \sigma_k | \log \rho_{q|z_{0:L}, A}  - \log  \rho_{q,  A} |\sigma_i \rangle \notag\\
&=& \langle \sigma_k | \log \rho_{q|z_{0:L}, A} \oplus \log \rho_{q|z_{0:L}, B} - \log  \rho_{q,  A} \oplus \log\rho_{q, B} |\sigma_i\rangle \notag
\end{eqnarray}
where we have used the fact that $\rho_{c,A} = \rho_{q,A}$ and $\rho_{c|z_{0:L}, A} = \rho_{q|z_{0:L}, A} $ for all $z_{0:L}$ and $\ket{\sigma_k} = \ket{k}$ for all $s_k$ in $A$,

We consider Case (2). Here $s_i \in A$ and $s_k \in B$, and thus $\ket{\sigma_k}$ is always orthogonal to $\ket{\sigma_i}$. The direct sum structure then implies 
\begin{eqnarray}
\langle i | \log\rho_{c|z_{0:L}, A} \oplus \log\rho_{c|z_{0:L}, B}  - \log \rho_{c,A} \oplus \log \rho_{c,B} |k\rangle  &=&0  \\
&=& \langle \sigma_k | \log \rho_{q|z_{0:L}, A} \oplus \log \rho_{q|z_{0:L}, B} - \log  \rho_{q,  A} \oplus \log\rho_{q, B} |\sigma_i\rangle \notag
\end{eqnarray}

Finally we consider Case (3), here $s_i, s_k \in B$
\begin{eqnarray}
\langle i | \log\rho_{c|z_{0:L}, A} \oplus \log\rho_{c|z_{0:L}, B}  - \log \rho_{c,A} \oplus \log \rho_{c,B} |k\rangle   & = & (\log P(s_i |z_{0:L}) - \log P(s_i)) \delta_{ik} \\&=& 0 \notag \\ &=&  \langle \sigma_k | \log \rho_{q|z_{0:L}, B}  - \log  \rho_{q,  B} |\sigma_i \rangle \notag \\
&=& \langle \sigma_k | \log \rho_{q|z_{0:L}, A} \oplus \log \rho_{q|z_{0:L}, B} - \log  \rho_{q,  A} \oplus \log\rho_{q, B} |\sigma_i\rangle \notag
\end{eqnarray}
where we have used the  definition of $B =\{s_i \,\, |  \,\, P(s_i) = P(s_i | z_{0:L}) \textrm{ for al } z_{0:L} \textrm{  and  } s_i \notin A\}$ and the fact that  $\rho_{c,B}  = \rho_{c|z_{0:L}, B}$ and $\rho_{q,B}  = \rho_{q|z_{0:L}, B}$ for all $z_{0:L}$.

Thus for all $s_i, s_k$ and $z_{0:L}$ we have equality in Eq. \eqref{eq:breakdownofrusaki1}. It follows from the results in the proof of the forward direction that $w^{(L)}_c - w_q^{(L)} = 0$.

 This concludes our proof of Result 2.

\section{Work cost of responding online}\label{sec:appendixrealtime}

We also analyse the gap between the work cost of responding to inputs one at a time, and the work cost  per output when generating $L$ outputs at a time in the limit of large $L$,  i.e. $ w_{\mathrm{onl}} = w_c^{(1)} - \lim_{L \rightarrow \infty} w^{(L)}_c$.

 To do this we make use of some facts about the Kolmogorov Sinai entropy rate. In particular we use the result that for any general stochastic process where the future and past are governed by random variables $\future{X} = X_0X_1\ldots$ and $\past{X} =\ldots X_{-1}$ respectively, the Kolmogorov Sinai entropy rate of the stochastic process captures the intrinsic unpredictability in the process $h_x = H(X_0 | \past{X})$ -- i.e., the extent to which the next symbol cannot be predicted even given knowledge of the entire past. In the limit of large $L$ the entropy of a block of $L$ symbols of the pattern approaches
\begin{equation}
\lim_{L\rightarrow \infty} H(X_{0:L}) = \lim_{L\rightarrow \infty}( I(\past{X}; \future{X}) + L h_x).
\end{equation}
where $I(\past{X}; \future{X})$ is the mutual information between past and future \cite{crutchfield2003regularities}. Thus we find  $h_x = \lim_{L\rightarrow \infty} H(X_{0:L})/L$. When the process is i.i.d.~the equality is achieved for every $L$.

When the input driving is i.i.d., the Kolmogorov Sinai entropy of the joint input-output process $h_z = \lim_{L\rightarrow \infty} H(Z_{0:L})/L$ can also be re-expressed in terms of the $\epsilon$-transducer's internal state as $h_z = H(Z_0| S_0)$  \cite{barnett2015computational}. We can thus re-express the cost of online response  as

\begin{eqnarray}
w_{\mathrm{onl}}/k T \ln 2 & =&  I(Z_0; S_1) - H(Y_0| X_0)  \notag \\
&& - \lim_{L\rightarrow \infty} \frac{I(Z_{0:L}; S_L)- H(Y_{0:L}| X_{0:L})}{L}  \notag \\
&=& I(Z_0; S_1) - (H(Y_0, X_0) -H(X_0))  \notag \\
&& + \lim_{L\rightarrow \infty} \frac{ H(Y_{0:L} X_{0:L}) - H(X_{0:L})}{L}  \notag \\
&=& I(Z_0; S_1) - H(Z_0) + H(X_0) + h_z - h_x \notag \\
&= & I(Z_0; S_1) - (H(Z_0) - H(Z_0 | S_0)) \notag \\ && + (H(X_0) - H(X_0 | \past{X}))  \notag \\
&=& I(Z_0; S_1) - I(Z_0; S_0)
\end{eqnarray}
where in the last line we used the i.i.d. nature of the input process. This result corroborates information ratchet results where it is known that when extracting from patterns there is an additional modularity cost for processing different parts of the tape piecemeal \cite{boyd2018thermodynamics}.

\section{Case study of quantum scaling advantage in work costs}\label{sec:appendixgap}
Here we look at case studies where the thermodynamic advantage of quantum agents can grow without bound. We examine two different tasks and in each case we look at how the gap between quantum and classical work costs scales with parameters in the desired response behaviour.

\subsection{Brownian Motion on a ring}
Here we consider a family of processes $\{\mathcal{P}_{\Delta I}\}$ which approaches the behaviour of a particle diffusing on a ring in the limit where  $\Delta I \rightarrow 0$ \cite{garner2017provably}. Specifically  we associate the points on the circumference of the ring with real numbers in $[0,1)$.  At precision $\Delta I$ we coarse gaining the circumference of this circle into bins of size $\Delta I$. In the continuum limit (when the interval size $\Delta I \rightarrow 0$) we represent the particle's position by a real number in this interval. At each time-step the particle then evolves according to random walk, while outputting the location it lands in. This generates a sequence of real numbers governed by a Brownian motion dynamic, such that
\begin{equation}\label{eq:brownianmotion}
y_{t+1} = \textrm{Frac}[{ y_t + d}]
\end{equation}
such that $d$ is drawn from some distribution $G_{\mu,\sigma}(d) = (\sigma)^{-1} (2\pi)^{-\frac{1}{2}}\exp\left(-\frac{(d-\mu)^2}{2\sigma^2}\right)$,
and $\textrm{Frac}[a] = a - \lfloor a \rfloor$ where $\lfloor \cdot \rfloor$ is the floor function (e.g. $\textrm{Frac}[3.43] = 0.43$). This ensures the ring gets mapped back to itself under the diffusion process. In particular we set $\mu = 0$, and $\sigma \ll 1$ so that the particle diffuses slowly around the ring.

For finite $\Delta I$ we assume the ring gets broken into $N = \lfloor 1/\Delta I\rfloor$ segments labeled $k \in \{0,\ldots N-1\}$. This corresponds to keeping only a finite number of digits of precision in our expression for the particle's current location. Now we have a discrete output alphabet $y_t \in \{0,\ldots, N-1\}$, corresponding to which of the $N$-intervals the particle lands in. A particle starting in location $k $ then transitions to location $j$  with probability $P_{kj} = G_{0, \sigma}(d) = (\sigma)^{-1} (2\pi)^{-\frac{1}{2}}\exp\left(-\frac{d^2}{2\sigma^2}\right)$ for $d = |j - k | \textrm{ mod } N$.

We can now formally define a strategy $\mathcal{P}_{\Delta I}$ that depends on two possible inputs, for each $\Delta I$. When the agent receives input $x_t = 0$ it must evolve as described above while emitting the label of the interval it lands in.  When it receives input $x_t = 1$ it first jumps $\pi$-radians then evolves as described above.

Since the dynamics in Eq. \eqref{eq:brownianmotion} are Markovian, such an agent's internal states are in one-to-one correspondence with the last output $y_t \in \{0,\ldots, N-1\}$. A classical agent thus has $N$ internal states associated with  computational basis elements, $s_i = \ket{i}\bra{i}$ for $i\in \{0,\ldots, N-1\}$. Due to the symmetry in the system, all $N$ states occur with equal probability $P(s_i) = 1/N$ in the steady state. 

Its quantum counterpart is implemented by quantum causal states identified by $\psi_q : s_i \rightarrow \ket{\sigma_i}\bra{\sigma_i}$, where
\begin{equation}\label{eq_csbrownian}
\ket{\sigma_i} = \sum_{i}\sqrt{P_{ik}}\ket{k}
\end{equation}

\begin{figure}[tbh!]
\centering
\includegraphics[scale=0.60]{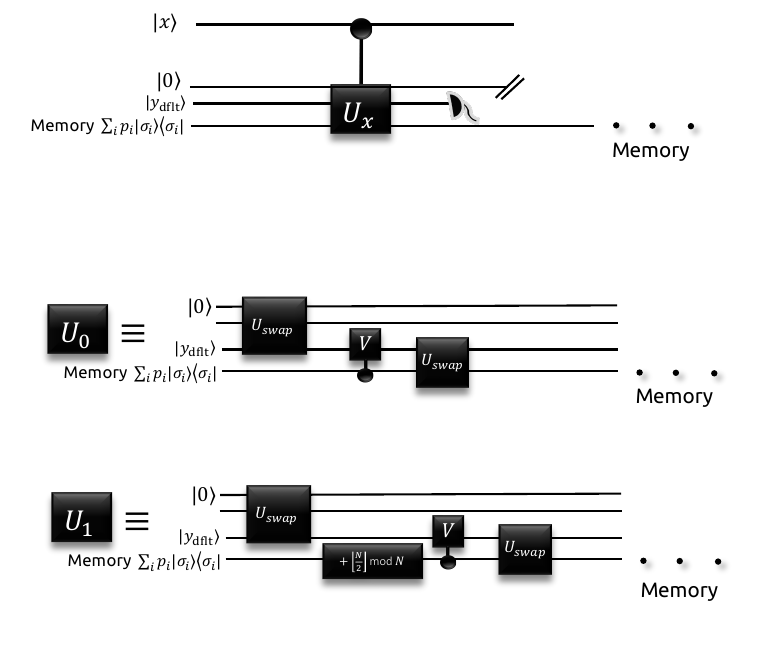}
\caption{\label{fig:brownian} A figure depicting the operation of the agent for the Gaussian random walk using the quantum causal states in  Eq. \eqref{eq_csbrownian}. Here the unitary $C-U = \ket{x}\bra{x}_c \otimes U_x$ is being controlled on the input register encoding the input stimuli $x_t$ at each point in time $t$. Meanwhile the two bottom panels show the specific conditional operations $U_x$.  In particular unitary $U_0$ (corresponding to input $x =0$), first uses a swap-gate, $U_{\textrm{swap}}$, to swap the battery register with the default state of the thermal tape $\ket{y_{\textrm{init}}}$. Afterwards the agent implements the transformation $\ket{0}\ket{\sigma_i} \rightarrow \sum_{ij} \sqrt{P_{ij}} \ket{j} \ket{\sigma_j}$  by harnessing a gate $ U_{\textrm{swap}} C-V = U_{\textrm{swap}} \ket{k}\bra{k} \otimes V_k$ such that $\ket{\sigma_k} = V_k\ket{0}$. Meanwhile  upon input $x = 1$ the agent applies $U_1$, which first increments the memory register basis states $\ket{k} \rightarrow \ket{k +\lfloor N/2\rfloor \textrm{ mod } N}$. The effect of this gate on the memory register is to map every memory state to its counterpart on the diametrically opposite side of the circle, i.e. $\ket{\sigma_i} \rightarrow \sum \sqrt{P_{ij}} \ket{j + \lfloor N/2\rfloor \textrm{ mod } N} = \ket{ \sigma_{(i + \lfloor N/2 \rfloor \textrm{ mod } N)}}$. Afterwards the rest of the gate proceeds identically to the $x= 0 $ case. Note that while we have a measurement operator in this circuit we assume this measurement is implemented by a von Neumann measurement which uses extra ancillary qubits borrowed from the battery register to realise (we do not explicitly depict this above; the image of a detector represents this von Neumann measurement circuit element). All qubits in the battery register  as well as the extra ancillary  battery qubit used for the von Neumann measurement must subsequently be reset following the protocols presented in preceding sections.}
\end{figure}

The unitary dynamics that allow it to generate appropriate future output responses for every possible input are then given in Fig. \ref{fig:brownian}.

To compute the thermodynamic advantage of the quantum agent over its classical counterpart, we evaluate 
\begin{equation}
\Delta W^{(L)} =   W^{(L)}_c - W^{(L)}_q  = k T \ln 2 \left[I(Z_{0:L}; S_L) - I(Z_{0:L}; M_L)\right]
\end{equation}
As the strategy is Markovian $H(S_1 |Z_{0:L}) = 0$ for all $L\ge 1$. It follows that the difference in the quantum and classical work cost is the same for all $L$ and thus we can simplify the above to
\begin{eqnarray}
\Delta W^{(L)}& =&  k T \ln 2 \left[I(Z_{0:L}; S_L) - I(Z_{0:L}; M_L)\right] \notag \\
& =& k T \ln 2 \left[H(S_L) - H(S_L | Z_{0:L})
- H(M_L) + H(M_L | Z_{0:L})\right]\notag \\ 
& =& k T \ln 2 \left[H(S_0) 
- H(M_0) \right]
\end{eqnarray}
where in the last line we have invoked stationarity $H(S_L) = H(S_0)$.  Furthermore we inherit a closed form bounds for both $H(S_0) = \log_2 N$ and $H(M_0) \le \left(\frac{1}{2\ln 2} - (1+4\sqrt{2\pi}\sigma)\log_2 2\sqrt{2\pi}\sigma\right)$ -- i.e. the quantum memory cost is bounded and finite for all $N$ but classical memory cost diverges logarithmically with $N$ \cite{garner2017provably}. This leads to a lower bound on the difference between the quantum and classical work cost 
\begin{eqnarray}
\Delta W^{(L)} &\ge& kT \ln 2 \left[ \log_2 N - \left(\frac{1}{2\ln 2} - (1+4\sqrt{2\pi}\sigma)\log_2 2\sqrt{2\pi}\sigma\right)\right] \notag \\ 
& = &  kT \ln 2\left[ \log_2 \left\lfloor \frac{1}{\Delta  I}\right\rfloor - \left(\frac{1}{2\ln 2} - (1+4\sqrt{2\pi}\sigma)\log_2 2\sqrt{2\pi}\sigma\right)\right]
\end{eqnarray}
We can see that for any $L$, this diverges as interval size $\Delta I \rightarrow 0$.

\subsection{Time tracking} \label{sec:time_tracking}
Here we consider a family of processes $\{\mathcal{P}_{\Delta t}\}$, that approaches the behaviour of a continuous time stochastic reset clock (i.e. a stochastic stopwatch) as $\Delta t \rightarrow 0$. Specifically, first consider a continuous time process defined by a stochastic clock that at any time can choose either to tick (representing output action $y_t = 1$) or stay silent (representing output action $y_t = 0$)~\cite{marzen2015, elliott2018superior, elliott2020extreme}. The clock is stochastic, such that the period between ticks is not fixed. Instead the clock has a survival probability $\Phi(T)$ of having a time-interval of at least $T$ seconds between neighbouring ticks. In addition, the clock is required to continually monitor for inputs. If the input is null ($x_t = 0$), it proceeds normally. However, should it receive $x_t = 1$ at any time $t$, the clock must immediately tick (i.e., emit $y_t = 1$ and reset). We refer to such an object as a stochastic reset clock~\cite{elliott2022quantum}. We consider a specific family of survival probabilities described by $\Phi(T) = p e^{-\gamma_0 T} + (1-p) e^{-\gamma_1 T}$ for some parameters $\{\gamma_0,\gamma_1\}$.

Now, consider a family of strategies $\{\mathcal{P}_{\Delta t}\}$ which represent a temporal coarse-graining of this behavior.
Each $\mathcal{P}_{\Delta t}$ represent a required response strategy for an agent that receives an input every $\Delta t $ seconds, such that as $\Delta t \rightarrow 0$, we approach the limit of the continuous time reset clock. Meanwhile, the agent is allowed to respond in a way that is only partially online. While the agent receives an input every $\Delta t$ seconds, it is allowed to collect questions over a fixed time period $\tau$, and respond to $\tau/\Delta t$ questions at a time. This is equivalent to setting $L = \tau/\Delta t$ in our framework.

We examine the energetic cost of realizing such an agent in the quasi continuous-time limit where $\Delta t \rightarrow 0$. In this setting we are interested in the amount of energy saved per unit time by the quantum agent, i.e,

 \begin{eqnarray}
\lim_{\Delta t \rightarrow 0} \frac{\Delta W^{(\tau/\Delta t)}}{\tau} = \lim_{\Delta t \rightarrow 0} (W^{(\tau/\Delta t)}_c -  W^{(\tau/\Delta t)}_q)/ \tau. 
\end{eqnarray} 
Here we show this quantity can diverge for fixed $\tau$. 

We start by defining the family of strategies $\{\mathcal{P}_{\Delta t}\}$ that represent a temporal coarse-graining of the continous-time reset clock. Each  $\mathcal{P}_{\Delta t}$ describes a strategy operating in discrete time, at each time-step $t\in \mathbb{Z}$ the input is a binary number $x_t \in \{0,1\}$ and the agent is required to decide either to tick ($y_t = 1$) or remain silent ($y_t = 0$). To faithfully execute $\mathcal{P}_{\Delta t}$, an agent must output $y_t = 1$ whenever $x_t = 1$. Otherwise on input $x_t = 0$, their choice of ticking should generally be dependent on the number of time-steps since the last tick, such that the survival probability of having at least $n$-zeros since the last tick follows the distribution $\Phi(n) = p\Gamma_0^n + (1-p) \Gamma_1^n$, with $\Gamma_i = e^{-\gamma_i \Delta t}$. In the limit $\Delta t \rightarrow 0$, an agent executing such a strategy resembles that of the stochastic reset clock. Note that under these circumstances the agent's strategy $\mathcal{P}_{\Delta t}$ is explicitly changing as a function of $\Delta t$.

\begin{figure*}[tbh!]
\centering
\includegraphics[scale=0.80]{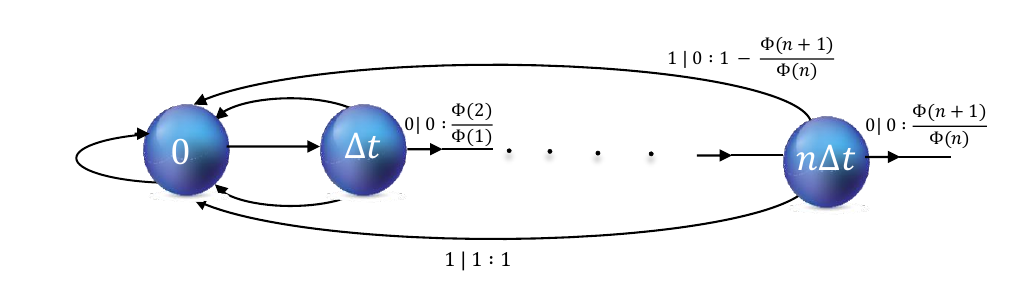}
\caption{\label{fig2} A figure depicting the  input-output relations $y|x: P(s_j, y | x , s_i)$ required to execute a stochastic reset clock as described in the text. These input-output relations are associated with edge labels. Meanwhile the nodes are the causal states of the model, where $s_i$ is being labeled $i\Delta t$ as it is associated with `surviving' $i$ time steps since the last tick event.}
\end{figure*}

\begin{figure}[tbh!]
\centering
\includegraphics[scale=0.60]{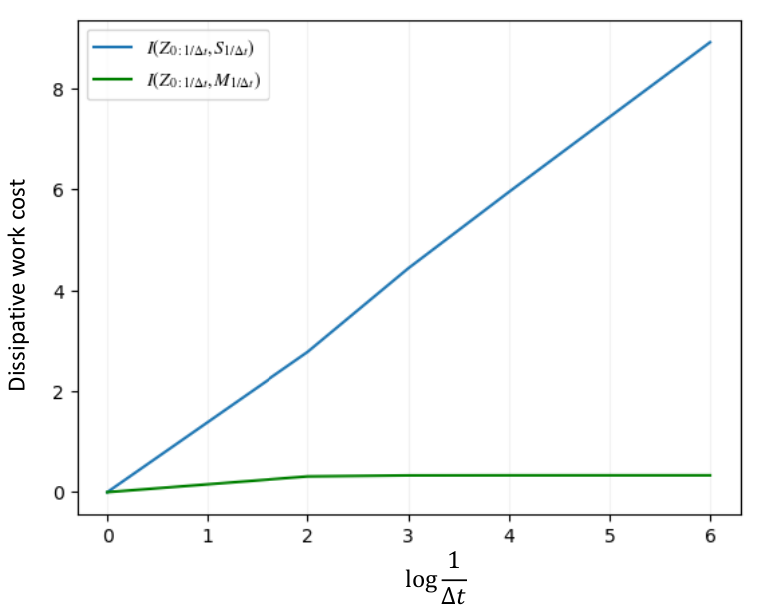}
\caption{\label{fig_scaling_plot} A figure depicting numerical estimates to the classical, represented by a blue line, (and quantum represented by the green line) dissipative work cost $kT\ln 2/\tau I(Z_{0:L}, S_L)$ for $L = \tau/\Delta t$, plotted against $\log(1/\Delta t)$. For convenience we choose free parameters  (i.e. $T,\tau$) which allow us to set the pre-factor $kT\ln 2/\tau$  to 1. To numerically estimate the classical dissipative work cost we have also truncate sums such as the one in Eq.~\eqref{eq:condition_entropy} at a maximal value of $L = 1/\Delta t$ and merged remaining states above this limit into one consolidated state (this cutoff value was chosen as the resulting approximation introduced minimal artefacts on the plotted lines). 
} \end{figure}

Any classical agent executing this strategy must track how many zeroes have been emitted since the last 1 emitted \cite{marzen2015, elliott2018superior, elliott2022quantum}. Thus the classically-optimal agent aims to store this information and nothing more. Their corresponding memory states will then be in one-to-one correspondence with the number of $0$s since the last $y= 1$ `tick event' -- i.e., the classical agent's  encoding function $\epsilon(\past{z}) = s_i$ if and only if $ y_{-i-1:0} = 1 0\ldots 0$. This leads to the construction in Fig.~\ref{fig2}. 

A quantum agent exhibiting this behaviour can be implemented using a single qubit of memory. Indeed we can adapt results from a recent analysis of a stochastic clock (which exhibits the desired $x_t =0$ behaviour \cite{elliott2020extreme, elliott2021quantum, wu2023implementing}), to arrive at a quantum memory state encoding $\epsilon_q = \psi_q\cdot \epsilon$ where $\psi_q : s_n \rightarrow \ket{\sigma_n}\bra{\sigma_n}$ for
\begin{eqnarray}\label{eq:quantumagentstates}
\ket{\sigma_n} &=& \ket{\varsigma_0} \otimes \ket{\varsigma_n},  \\
\ket{\varsigma_n}&=& \frac{\sqrt{p\Gamma_0^n}}{\sqrt{\Phi(n)}} \ket{h_0} + i\frac{\sqrt{\overline{p}\Gamma_1^n}}{\sqrt{\Phi(n)}} \ket{h_1}, \notag
\end{eqnarray}

such that $\Phi(n) = p\Gamma_0^n + \overline{p}\Gamma_1^n$ , $\overline{p} = 1-p$  and $\Gamma_i = e^{-\gamma_i \Delta_t}$; meanwhile $\ket{h_0} = \ket{0}$ and $\ket{h_1} = g\ket{0} + \sqrt{1-g^2} \ket{1}$ for
\begin{equation}
g = \frac{\sqrt{(1-\Gamma_0) (1- \Gamma_1)}}{1-\sqrt{\Gamma_0\Gamma_1}}.
\end{equation}
 
\begin{figure}[tbh!]
\centering
\includegraphics[scale=0.60]{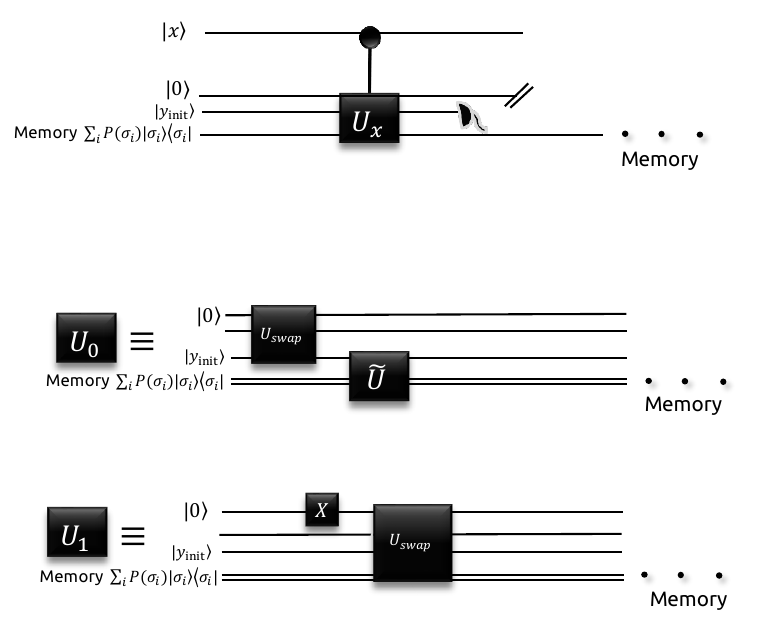}
\caption{\label{fig3} A figure depicting the operation of the resettable quantum agent whose internal states are described by Eq.~\eqref{eq:quantumagentstates}. Here the unitary is being controlled on the input register encoding the input stimuli $x_t$ at each point in time $t$. Meanwhile the two bottom panels show the specific conditional operations $U_x$ implemented via a control gate $C-U = \ket{x}\bra{x}_c \otimes U_x$. We associate the wires in this diagram with inputs $\ket{x}_c\ket{0}_{t1}\ket{0}_{t2} \ket{y_{\init}}_{t3} \ket{\sigma_n}_{t4,t5}$ where $c$ is the subspace of the control, and target wires $t1$ and $t2$ are the battery, $t3$ is associated with the output tape, and the memory is encoded in wires $t4, t5$ as $\ket{\sigma_n}_{t4,t5} = \ket{\varsigma_0}_{t4}\ket{\varsigma_n}_{t5}$. The form of $U_{x = 0} = \widetilde{U}_{t3,t5} U_{\textrm{swap}(t1,t3)}$ where $U_{\textrm{swap}(a,b)}$ swaps wires $a$ and $b$, meanwhile the form of $\widetilde{U}_{t3,t5}$ is given exactly in the supplementary materials of \cite{elliott2020extreme}. If a wire does not have a gate acting on it then we assume that wire evolves under the identity channel. Meanwhile $U_{x = 1} = U_{\textrm{swap}(t2,t5)} U_{\textrm{swap}(t1,t3)}X_{t1} $. Note that while we have a measurement operator in this circuit we assume this measurement is implemented by a von Neumann measurement which uses extra ancillary qubits borrowed from the battery register to realise (please note we do not explicitly depict this above. We use the image of a detector to represent this von Neumann measurement circuit element.). All qubits in the battery register represented by wires $t1, t2$, as well as the extra ancillary  battery qubit used for the von Neumann measurement must subsequently be reset following the protocols presented in preceding sections.
} \end{figure}

With these states we can build a quantum agent. For a detailed break down of the circuit (c.f.  Fig.~\ref{fig1} for this model) see Fig.~\ref{fig3}.

To evaluate the energetic expenditure of responding to the input, we assume the inputs follow an i.i.d.~distribution at each time step governed by a random variable $X_t = \Gamma_X \ket{0}\bra{0} + (1-\Gamma_x) \ket{1}\bra{1}$, where we set $\Gamma_x = e^{-\gamma_x \Delta t }$. Under these circumstances we find that the probability of an agent being in causal state $s_n$ is $\pi_n = \mu \widetilde{\Phi}(n)$, where
\begin{equation}
\widetilde{\Phi}(n) = p \widetilde{\Gamma}_0^n  + \overline{p} \widetilde{\Gamma}_1^n
\end{equation}
for  $\widetilde{\Gamma}_i = \Gamma_X \Gamma_i$ \cite{elliott2020extreme, marzen2015, elliott2022quantum}. Meanwhile $\mu^{-1} = \sum_n \widetilde{\Phi}(n) = \sum_{n } p\widetilde{\Gamma}_0^n + \overline{p} \widetilde{\Gamma}_1^n$, is a sum of two  geometric progressions with closed form expression
\begin{equation}
\mu = \frac{(1-\widetilde{\Gamma}_0)(1- \widetilde{\Gamma}_1)}{p(1-\widetilde{\Gamma}_1) + \overline{p}(1-\widetilde{\Gamma}_0)}.
\end{equation}

The corresponding steady state for the classical agent is $\rho_c = \sum_{n} \mu \widetilde{\Phi}(n)  \ket{n}\bra{n}$.  Meanwhile the quantum agent's memory register is $\rho_M = \sum_{n} \mu \widetilde{\Phi}(n) \ket{\sigma_n}\bra{\sigma_n}$. In the limit of small $\Delta t $ these quantities approach probability density functions, such that $\pi_t \approx \mu \widetilde{\Phi}(t) \Delta t$ where $\widetilde{\Phi}(t) =  p e^{-(\gamma_0 +\gamma_x) t} + (1-p) e^{-(\gamma_1 + \gamma_x) t}$ and  $\mu^{-1} = \int_0^{\infty} \widetilde{\Phi}(t) dt $.

We now investigate how the classical-quantum work gap per unit time scales in the limit of responding in infinitesimal time $\Delta t \rightarrow 0$. To do this we place bounds on the quantity
\begin{equation}
(W^{(\tau/\Delta t)}_c -  W^{(\tau/\Delta t)}_q)/ \tau = kT \ln 2/ \tau [I(Z_{0:L}; S_L) - I(Z_{0:L}; M_L)]
\end{equation}
for each value of $\Delta t$  (corresponding to each value of $L = \tau /\Delta t$).

To do this observe that for any output response string $y_{0:L} \neq 0\ldots 0$ we have $H(S_L|y_{0:L}) =  0$. With out loss of generality we can assume $\widetilde{\Gamma}_1 \le \widetilde{\Gamma}_0$. It follows that
\begin{eqnarray}
H( S_L|Z_{0:L}) &=& P(z_{0:L} =(0\ldots 0,0\ldots 0)) H(S_{L}|z_{0:L} = (0\ldots 0, 0\ldots 0))  \label{eq:condition_entropy} \\ && + \sum_{y_{0:L} \neq 0\ldots 0} P(z_{0:L}) H(S_L|z_{0:L}) \notag \\   &=& P(z_{0:L} = (0\ldots 0,0\ldots 0)) H(S_L|z_{0:L}= (0\ldots 0,0\ldots 0)) \notag \\
& \le &  \widetilde{\Gamma}_0^L  H(S_{0}) \notag \\
& = & e^{-(\gamma_0 +\gamma_x) \tau} H(S_0) \notag
\end{eqnarray}
where we have used an upper bound on the survival probability, as well as stationarity of the agent's internal state to simplify the second-to-last line.  In particular, we have used the fact that the output string $y_{0:L} = 0\ldots 0$ can only be observed if the input driving string is $x_{0:L} =0 \ldots 0 $. Meanwhile, for our chosen i.i.d.~input driving  we have $P(x_{0:L} = 0\ldots 0) = \Gamma_X^L = e^{-\gamma_x\Delta t L} = e^{-\gamma_x \tau}$. Finally, the survival probability directly bounds the probability of seeing $L$ contiguous zero outputs, conditioned on getting input $x_{0:L} = 0\ldots 0$. We adapt this formula to get the upper bound $P(y_{0:L} = 0\ldots 0| x_{0:L} = 0\ldots 0) \le \Gamma_0^L = e^{-\gamma_0 \tau}$.

These results allow us to bound $I(S_L; Z_{0:L}) \ge (1-  e^{-(\gamma_0 +\gamma_x) \tau}) H (S_0)$. Meanwhile there exists a closed form approximation to $C_{\mu} = H(S_0)$ in the limit of small $\Delta t $  \cite{crutchfield2015time, marzen2015, elliott2020extreme},
\begin{equation}
C_{\mu} \approx \log_2\left(\frac{1}{\mu \Delta t}\right) - \mu^{-1} \int_0^\infty \widetilde{\Phi}(t) \log_2 \left(\widetilde{\Phi}(t) \right) dt
\end{equation}
This expression scales $C_{\mu} \sim \log_2\left( \frac{1}{\Delta t} \right)$ as $\Delta t \rightarrow 0$.

In addition we can trivially upper bound $I(M_L; Z_{0:L} ) \le H(M_0) \le H_{\textrm{max}}(M_0) \le 2$, due to the capability to realise the quantum agent with at most 2 qubits of memory in Eq.~\eqref{eq:quantumagentstates}~\cite{elliott2020extreme}\footnote{Note that strictly only one qubit of memory is required to realise the quantum agent, as the $\{\ket{\sigma_n}\}$ span only a 2-dimensional Hilbert space. The bound can therefore be tightened to $H_\mathrm{max}(M_0)\leq1$, though this does not materially affect our result.}. 

It flows directly from the above that $I(M_L, Z_{0:L}) \le H(M_L) \le 2$ and is thus finite and bounded for any value of $\Delta t$. Meanwhile the classical dissipative work cost depends directly on $I(S_L, Z_{0:L})  = H(S_L) - H(S_L |Z_{0:L}) \ge (1 - e^{-(\gamma_0 + \gamma_1) \tau}) H(S_0) \sim ( 1-  e^{-(\gamma_0 + \gamma_1) \tau})\log_2\left( \frac{1}{\Delta t} \right)$ which diverges with $\log_2\left( \frac{1}{ \Delta t}\right)$  as $\Delta t \rightarrow 0$.

 Using the above results in conjunction we can express:

\begin{eqnarray}\label{eq:contunuumlimitworkclock}
\lim_{\Delta t \rightarrow 0} (W^{(\tau/\Delta t)}_c -  W^{(\tau/\Delta t)}_q)/ \tau & = & \lim_{\Delta t \rightarrow 0} \frac{kT \ln 2}{\tau} [I(S_L; Z_{0:L})  - I(M_L; Z_{0:L})]\notag \\
& \ge&  \lim_{\Delta t \rightarrow 0}\frac{ kT \ln 2}{\tau} [(1-\Gamma_X^Le^{-\gamma_0 \tau}) H(S_0) - 2]\notag \\
& \approx & \lim_{\Delta t \rightarrow 0} \frac{kT \ln 2 }{\tau} (1-e^{-(\gamma_0 +\gamma_x)\tau})\log_2\left(\frac{ 1}{\Delta t}\right)
\end{eqnarray}

We plot numerical estimates for the classical disipative work component of executing this task $kT\ln 2/\tau I(Z_{0: L}, S_{L})$ (and its quantum counterpart) against $\log (1/\Delta t)$ in Fig.~\ref{fig_scaling_plot} . As the $x$-axis of this plot is log scale, the resulting linear relationship between $\log_2(1/\Delta t )$ and the classical disipative work cost is indicative of an exponential divergence in classical disipative work cost per unit time with $1/\Delta t$.

\section{Thermodynamic benefits of using a higher dimensional memory}

Classically we can identify the thermodynamically optimal agent construction, and show that it corresponds with the classical model which has the lowest memory dimension (and lowest Shannon entropy) -- the strategies' $\epsilon$-transducer. However,  we have no current way of finding a thermodynamically optimal (or memory optimal) quantum model. To identify the optimal quantum model we need to optimize the encoding of classical memory states into quantum counterparts, minimizing the entropy of the ensemble while keeping different states sufficiently discriminable to allow future output responses to be generated by a completely positive trace preserving map. 

\begin{figure*}[tbh!]
\centering
\includegraphics[scale=0.80]{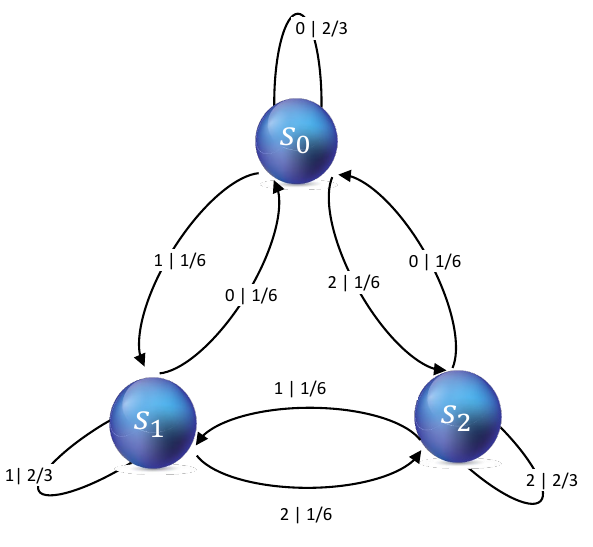}
\caption{\label{fig3tatemachine} An input independent process which features 3 causal states from \cite{loomis2019strong}. Edges are labeled by $y|P(s_j, y| s_i)$.}\end{figure*}

This opens some interesting questions, such as how can we increase the degree of thermodynamic advantage?  Can we improve thermodynamic performance by using a quantum agent with a higher dimensional memory?  Indeed for the simple case where the behaviour is independent of the input, it has already been established that the i.i.d. memory cost of an agent can go down as we increase dimension of the Hilbert space spanned by its memory states \cite{loomis2019strong, liu2019optimal}. We reproduce one such case in Fig.  \ref{fig3tatemachine}. We highlight that this example simultaneously proves the thermodynamic cost can also be reduced by increasing the dimensionality of the quantum memory.

To see this note that this process has two potential models with respective internal states $S_1 = \{\ket{m_0},\ket{m_1}, \ket{m_2}\}$, and $S_2 = \{\ket{n_0},\ket{n_1}, \ket{n_2}\}$, described by:
\begin{align}
&\ket{m_0} = \ket{0} \qquad & \ket{n_0} = \sqrt{\frac{2}{3}} \ket{0} + \frac{1}{\sqrt{6}}(\ket{1} + \ket{2})\notag \\ 
&\ket{m_1} = \frac{1}{2} \ket{0} + \frac{\sqrt{3}}{2} \ket{1} \qquad &  \ket{n_1} = \sqrt{\frac{2}{3}} \ket{1} + \frac{1}{\sqrt{6}}(\ket{0} + \ket{2})\notag \\ 
&\ket{m_2} = \frac{1}{2} \ket{0} - \frac{\sqrt{3}}{2} \ket{1} \qquad &  \ket{n_2} = \sqrt{\frac{2}{3}} \ket{2} + \frac{1}{\sqrt{6}}(\ket{0} + \ket{1}) \notag \\ 
&H\left(\sum_i P(m_i) \ket{m_i}\bra{m_i}\right) = 1\qquad  & H\left(\sum_i P(n_i) \ket{n_i}\bra{n_i}\right) = 0.61 
\end{align}
Since the process is Markovian, we automatically have $I(Z_0, M_1) = H(M_1)$. We see from the form of Eq. \eqref{eq:iidquantumcost} that the work cost 
\begin{equation}
W^{(1)}_q/k T \ln 2 =  \hdef - H(Y_0|X_0)  + I(Z_0; M_1).
\end{equation}
Thus the model with the lowest value of $I(Z_0, M_1) = H(M_1)$ will be the most thermodynamically efficient. While the model on the right based on $S_2 = \{\ket{n_0},\ket{n_1}, \ket{n_2}\}$ has a higher memory dimensionality, it nonetheless has a lower $H(M_1)$ and thus is the more thermodynamically efficient model. 

This establishes that increasing the memory dimension of the quantum model can in fact improve its thermodynamic performance.

\end{document}